\documentclass[a4paper, accepted=2026-05-14]{quantumarticle}
\pdfoutput=1

\usepackage{graphicx} 
\usepackage{dcolumn}
\usepackage{bm}
\usepackage{amsmath,latexsym,tabularx}
\usepackage{multirow}
\usepackage[export]{adjustbox}
\usepackage{hyperref}
\usepackage{upgreek}
\usepackage{textgreek}
\hypersetup{
  colorlinks=true,
  urlcolor=blue,
  linkcolor=blue,
  citecolor=blue
}

\usepackage{todonotes}
\usepackage[dvipsnames,svgnames,x11names]{xcolor}
\usepackage[marginparwidth=2.5cm]{geometry}
\usepackage[numbers,sort&compress]{natbib}
\usepackage[utf8]{inputenc}
\usepackage[english]{babel}

\usepackage{pythonhighlight} 

\newcommand{\affilUMass}[0]{\affiliation{University of Massachusetts Amherst, Dept. Electrical and Computer Engineering, Amherst, MA}}

\definecolor{codegreen}{rgb}{0,0.6,0}
\definecolor{codegray}{rgb}{0.5,0.5,0.5}
\definecolor{codepurple}{rgb}{0.58,0,0.82}
\definecolor{backcolour}{rgb}{0.97,0.97,0.97}
 
\lstdefinestyle{myPythonStyle}{
    backgroundcolor=\color{backcolour},   
    commentstyle=\color{codegreen},
    keywordstyle=\color{magenta},
    numberstyle=\tiny\color{codegray},
    stringstyle=\color{codepurple},
    basicstyle=\footnotesize,
    breakatwhitespace=false,         
    breaklines=true,                 
    captionpos=b,                    
    keepspaces=true,                 
    numbers=left,                    
    numbersep=5pt,                  
    showspaces=false,                
    showstringspaces=false,
    showtabs=false,                  
    tabsize=2,
    keywordstyle = [2]{\color{blue}},
    morekeywords = [2]{place_element_along_beam,beam,component,baseplate,ECDL,Rb_SAS,singlepass,doublepass,ecdl,rb_sas},
    keywordstyle = [3]{\color{orange}},
    morekeywords = [3]{beam_index,distance,angle,beam_obj,x,y,dx,dy,rotation,label,definition,position,dimensions},
}

\newcommand{\Sr}{$^{88}$Sr$^+$}

\begin{document}

\title{Qubit operations using a modular optical system engineered with PyOpticL: a code-to-CAD optical layout tool}

\author{Jacob Myers}
\affilUMass
\author{Christopher Caron}
\affilUMass
\author{Nishat Helaly}
\affilUMass
\author{Zhenyu Wei}
\affilUMass
\author{Justin Oh}
\affilUMass
\author{Zack Gotobed}
\affilUMass
\author{Kotaro Yabe}
\affilUMass
\author{ Robert J. Niffenegger}
\affilUMass

\date{5/14/2026}

\maketitle

\begin{abstract}
Complex optical design is hindered by conventional piecewise setup, which prevents modularization and therefore abstraction of subsystems at the circuit level. This limits multiple fields that require complex optics systems, including quantum computing with atoms and trapped ions, because their optical systems are not scalable.
We present an open-source Python library for optical layout (PyOpticL) which uses beam-path simulation and dynamic beam-path routing for quick and easy optical layout by placing optical elements along the beam path without a priori specification, enabling adaptive, path-based layouts with automatic routing and connectivity.
We use PyOpticL to create modular `drop-in' optical baseplates for common optical subsystems used in atomic and molecular optics (AMO) experiments including laser sources, frequency and intensity modulation, and locking to an atomic reference for stabilization. We demonstrate this minimal working example of a dynamic full laser system for strontium trapped ions by using it for laser cooling, qubit state detection, and over $99\%$ fidelity single-qubit gates with 3D printed baseplates.
This enables a new paradigm of design abstraction layers for engineering optical systems leveraging modular baseplates, as they can be used for any wavelength in the system and enables scaling up the underlying optical systems for quantum computers.
This new open-source hardware and software code-to-CAD library seeks to foster open-source collaborative hardware and systems design across numerous fields of research including AMO physics and quantum computing with neutral atoms and trapped ions.
\end{abstract}

\section{Introduction}

Careful optical design is critical for compact, high-performance optical layouts used in atomic and molecular optical (AMO)  experiments for fundamental physics and sensing applications, as well as quantum computers based on neutral atoms and trapped ions. 
Miniaturizing layouts to optical baseplates \cite{kulkarni2020ultrastable, zhang2022design} and micro-optics \cite{knappe2007microfabricated, maurice2020miniaturized, strangfeld2021prototype}
improves stability and performance, while simultaneously enabling rack mounting and portable operation. 
However, neither miniature baseplates nor micro-optics are widely used due to the difficulty of customizing and updating their designs using existing 3D CAD tools.

For electronics and VLSI (Very Large Scale Integration) design, tools like Cadence are critical enablers, which dynamically manage layouts while automatically routing connections between components.
VLSI design tools also utilize hierarchical abstraction of levels so that the functional level of design is not burdened by the device level. 

In photonics, open source tools for layout like GDSFactory \cite{gdsfactory} have created an open source collaborative community that dramatically lowers the barrier to entry for advanced and large-scale photonic layout.
Despite the advantages of using CAD for optical circuit design, there is no tool that can dynamically position optical elements and route connections simultaneously. 
This means that optical circuit design can not abstract away these aspects of design at the device level (optical element), preventing the abstraction to higher levels of the optical system hierarchy and therefore scalable optical systems engineering. 

Here we present and demonstrate an open-source Python tool for optical layout, called `PyOpticL'\cite{pyopticl} (\href{https://github.com/UMassIonTrappers/PyOpticL}{https://github.com/UMassIonTrappers/PyOpticL}). 
This optical layout library is built on another open source software tool, FreeCAD \cite{freecad}, which is also based on Python, making them seamlessly compatible and easily expandable.

Building on top of another open source tool itself, PyOpticL seeks to emulate and complement other open source hardware movements in the AMO and quantum computing community, such as ARTIQ (M-Labs) \cite{kasprowicz2020artiq}, which has created a central open source hardware development hub for the precise electronic control hardware required for precision AMO and quantum computing experiments.
This enables collaborative hardware projects which can be developed and maintained by a large and growing community of researchers.

Critically, the modularization of several optical circuits for modulation and stabilization enables abstraction of these levels within the larger AMO apparatus, as these same optical baseplates can be applied to all of the wavelength-specific subsystems. 

Additionally, the script-based nature of the optical layouts designed in PyOpticL enables the hardware design process to be managed via version control similar to software development, enabling easy publishing of hardware designs and collaborative convergence on best-known methods across disciplines.

\section{Background}

\subsection{Ad Hoc Piecewise Optics}

Piecewise placement and alignment of optical elements is a tedious and slow process. 
This burden prohibitively restricts how frequently optical layouts are updated or revised, as it is often not possible to revise the position or angle of any single optical element without perturbing the entire downstream optical beam path.

While it is possible to isolate optical degrees of freedom of optical path branches, through the use of polarizing beam-splitters (PBS) etc., this still does not often translate into a modular layout, due to the underlying monolithic nature of mounting all individual optical elements piecewise to a single optical table.

Smaller optical breadboards offer a way to modularize optical subsystems, but in practice are seldom utilized for this purpose. Even on a breadboard, the subsystems are often too large to be modular, and these breadboards themselves must be still manually aligned, one optical element at a time, to a fixed grid.

Smaller half-inch optics have been used to build compact optical systems for AMO applications such as trapped ions \cite{duke,pogorelov2021compact,spivey2021high,chen2022stable}.
However, they are often still set up ad hoc upon a fixed grid similar to an optical table, such that precise, serial placement and alignment of all optical elements is still required.

To overcome these challenges, there have been many ``in-house" \cite{JayichUCSBdoublepass, duke, zhang2022design} and also proprietary \cite{aosense, AQTrowan} solutions to optical layout using 3D CAD software like SolidWorks, which allow users to precisely place optical elements and define a custom baseplate to mount them.
This dramatically aids in the precise positioning and angular alignment during the placement of optical elements upon the baseplate, making optical setup as easy as `paint by numbers'.

This drop-in alignment is especially critical for micro-optics \cite{knappe2007microfabricated} which use 3 mm optical elements and are placed without any adjustable mounts, preventing re-alignment after elements are placed.


\subsection{Dynamic Optical Routing}

Even with the advantages of custom baseplates designed in 3D CAD software, these designs are still static with no way to dynamically update the optical elements in the circuit nor the layout of the circuit, which must be tweaked/aligned manually within the CAD software. This means that substantial alignment time is still required \textit{within} the CAD environment, eroding the time-saving benefits of pre-designing the layout. 

\begin{figure}
\centering
\includegraphics[width=0.49\textwidth]{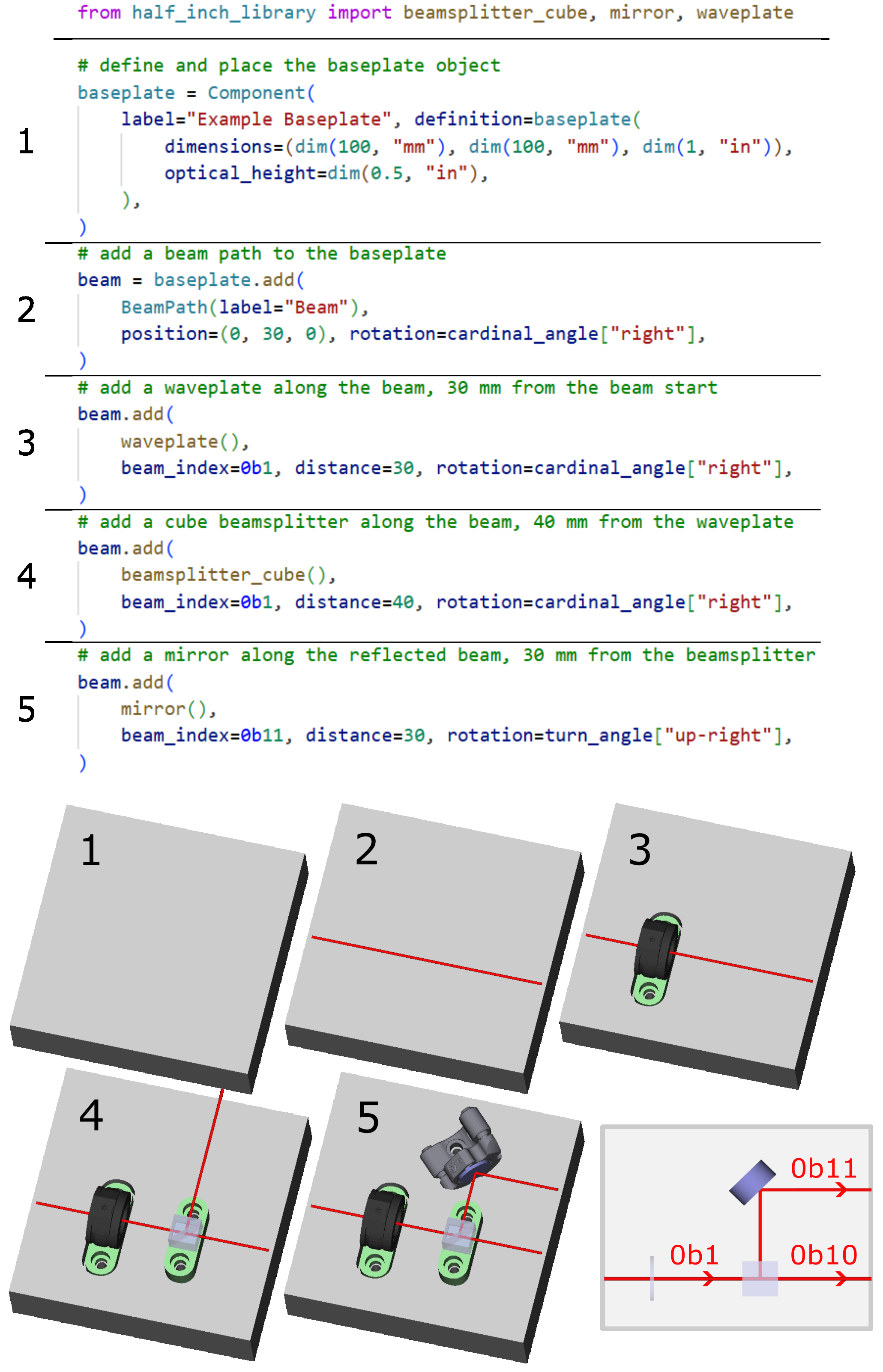}
\caption{
\textbf{Code-to-CAD system:}
An example of how a simple baseplate can be scripted. (1) First a baseplate object is initialized. (2) A beam can then be added to the baseplate at some input position and angle. (3-5) After this, all components can be place along the beam path by specifying the beam index (see Path-based layout creation), the distance from the previous component and the angle of the object.
}
\label{fig:Code-to-CAD}
\end{figure}

This also means that changing any single optical element in the design (or converting from 1/2-inch to micro optics) requires a total re-alignment of the layout within the CAD environment, limiting how broadly the static designs can be applied for various wavelengths etc. Even parameterizing the layout eventually becomes manual without beam tracking. Additionally, this beam alignment verification process is manual, as there is no ability within these CAD tools to track beam paths and verify beam alignment between optical elements.

Many of these issues stem from the lack of dynamic routing of the optical layout. All CAD software used for electronic and photonic circuit layout has some form of dynamic routing which aids in adjusting the layout without breaking connectivity within the circuit.
Here, we present a new architecture for optical layout based on our custom Python library (PyOpticL) which uses `beam-based routing', to automatically track the optical beam-path throughout the circuit and place optical elements \emph{along} the laser beam's path, as defined by a simple script. 
We also define a set of simple design rules (see SI for details) which enable a modular and scalable design for interchangeable parts within a complex optical system.

\begin{figure*}[]
\centering
\includegraphics[width=\textwidth]{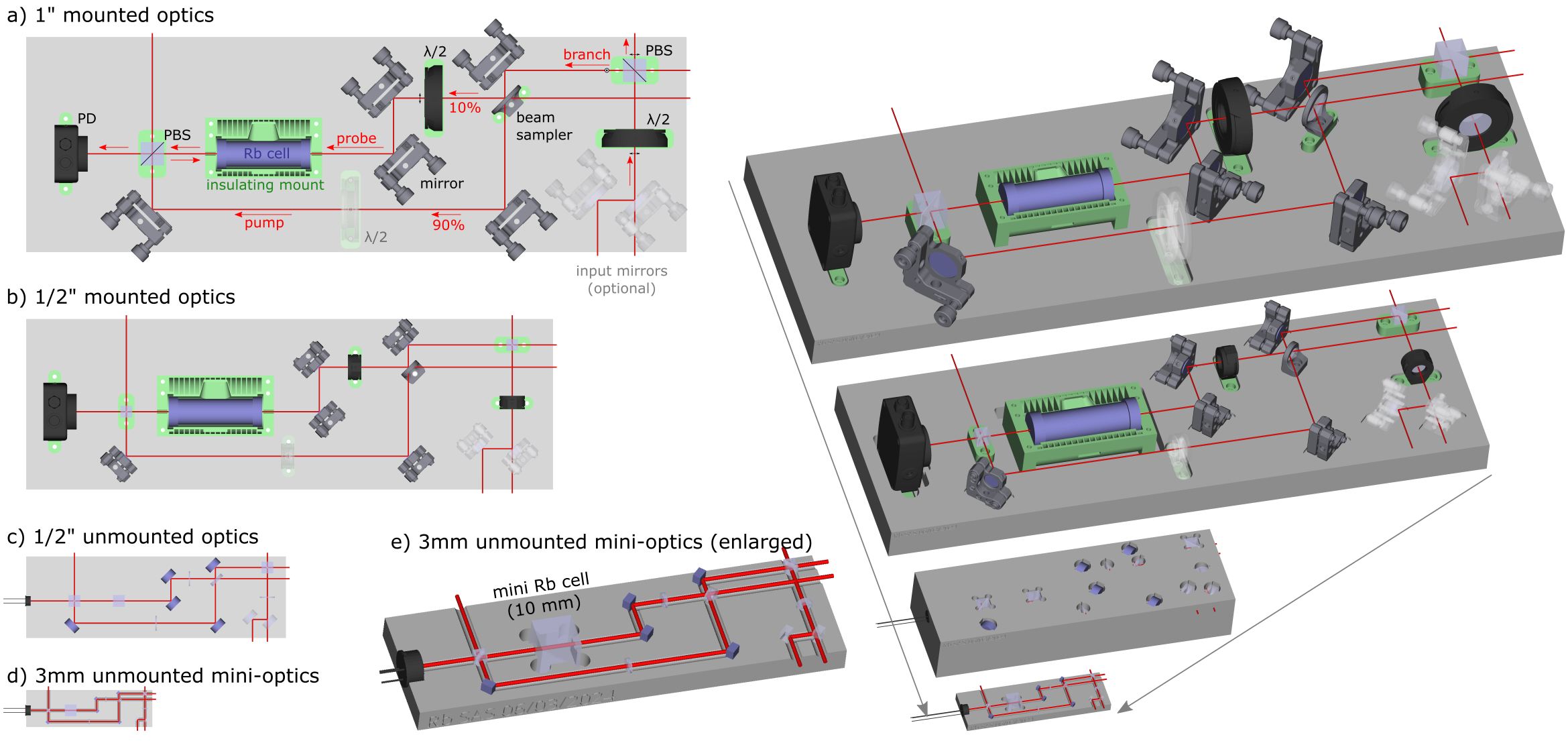}
\caption{
\textbf{Scalable layouts from full size to micro-optics:}
The same base optical layout code can be arbitrarily recompiled with different components for each optical element and at different scales. As an example, the same Saturated Absorption Spectroscopy layout script can be compiled at 4 different scales with 4 different sets of optics: (a) 1 inch optics, (b) 1/2 inch mounted optics, (c) `mount-free' 1/2 inch optics, (d) 3 mm micro-optics. The underlying layout is identical for all layouts, but the component definitions and distances are adjusted for each of the different scales, demonstrating the dynamic possibilities of code based CAD layout.
}
\label{fig:dynamic}
\end{figure*}

\section{PyOpticL}

\subsection{CAD Software}
The underlying CAD software upon which to build the PyOpticL library is critical. The most important requirements in choice of CAD software was robust and feature-rich scripting support coupled with an efficient graphical engine. For this reason, FreeCAD was chosen.

FreeCAD \cite{freecad,riegel2016freecad} provides both a fully featured graphical user interface and a Python-based scripting back-end. The Python back-end allows FreeCAD to be easily tailored for specialized applications through community-built add-ons, called workbenches, which add both scripting and GUI based functionality through importable Python modules. This python-based scripting allows for the creation of advanced systems such as are required for beam simulation and layout creation.

OpenSCAD was also considered for its script-only CAD approach. However, OpenSCAD's scripting is quite limited due to the fact that variables cannot be modified at runtime (e.g. $x= x+1$ is not possible) \cite{machado2019parametric}. This would make it impossible to implement the simulations and runtime automations facilitated by PyOpticL.

Additionally, non-free CAD options such as AutoCAD and SolidWorks do provide some scripting functionality, but they lack the flexibility required to create complex layout systems. AutoCAD scripts are only able to execute a subset of basic commands, limiting them to simple automations. In contrast, SolidWorks macros allow for the use of C\# or VB.NET frameworks, giving far more functionality. However, all scripts must be pre-compiled to DLL format before running, significantly impairing rapid iteration. Lastly, the use of non-free software would be contradictory to the motivation of this project to create an open-source software solution freely available to the entire community.

\subsection{Modular Design}
PyOpticL is built with a focus on modularity in all design elements. The fundamental unit in PyOpticL is the \textit{Layout} object. Simply put, the purpose of the Layout object is to allow relative positioning between all objects within a design. By calling the \textit{add} function, layouts can be nested together. This creates a grouping of objects with positions defined relative to the parent layout. We can then define a \textit{Component} object as a subclass of Layout, but with a physical geometry. As the Component object retains all functionality of the Layout class, components can be nested in layouts to create designs just the same as they could be nested into other components to create compound objects.

For example, baseplates, optics, and mounts are all Component objects. To create a design, you would define a baseplate, then add the appropriate components to it. To then create a full subsystem comprised of multiple baseplates (See Fig. \ref{fig:full}), those baseplate objects can simply be added to a top-level Layout object. In this way, PyOpticL is able to seamlessly handle the relative placement of all objects while still allowing unlimited nesting of layouts, components, etc.

\subsection{Beam Simulation}
In order to facilitate accurate modeling and dynamic routing for layout, a robust beam simulation system is required. PyOpticL uses a simplified, but feature-rich 3D beam-tracing system which allows for accurate placement and verification of optical layouts. To define how optical elements interact with the beam, we define classes of optical interfaces such as reflection, lens, waveplate, or diffraction. By linking physical component models to these interfaces, PyOpticL is able to track the path, power, polarization, and divergence of the beam as it passes through the layout. The extensibility of this interface system also allows new types of optical interactions to be added easily without modifying the underlying simulation framework. The overall goal of the implementation is to provide an efficient system for path-based layout creation while maintaining useful tracking features for design verification.

\subsection{Path-based Layout Creation}
The ability to sequentially predict the path of the beam as it interacts with optical elements allows a new paradigm of optical layout which sets new optics along the path of the beam dynamically, without using any fixed coordinates within the CAD system. 
To enable this, we define a \textit{BeamPath} object as a subclass of Layout with an adjusted \textit{add} function. Instead of utilizing absolute positioning relative to the parent, this add function allows an object's position to be defined by the beam on which it will be placed, the sub-beam index, and an additional constraint such as a distance from the last component or a single absolute x, y, or z coordinate.

The sub-beam index here refers to the way in which the beam path is segmented in the case of splits or diffraction. In order to facilitate placement along any arbitrarily branching path, a non-repeating beam indexing system is required. For this case, a binary tree system was used (Fig. \ref{fig:Code-to-CAD}). The initial beam is given index one, or `0b1' as a binary literal. When the beam passes through a splitter, a zero or one is appended to the binary index of the transmitted (0) and reflected (1) beams respectively. So in the case of the first split in a path, the transmitted beam would have index `0b10' and the reflected beam would have index `0b11'. In this way, not only are the beam indexes unique for any arbitrary path, but they remain highly readable due to the direct correlation to the binary literal and the path taken to arrive at that beam segment.

\begin{figure*}[]
\centering
\includegraphics[width=0.9\textwidth,height=\textheight,keepaspectratio]{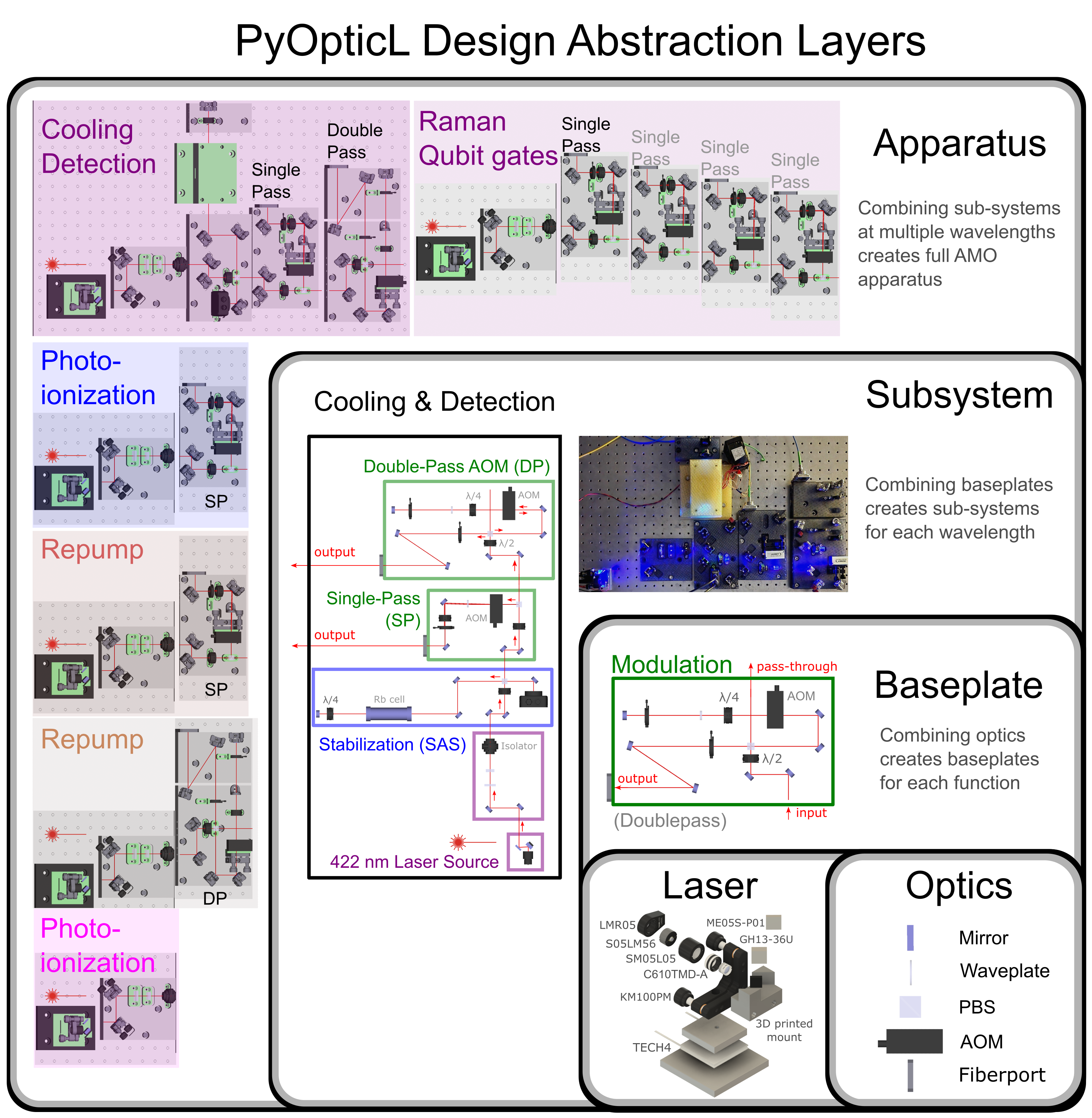}
\caption{
\textbf{Hierarchical layers of design abstraction that are enabled by PyOpticL applied to a strontium trapped ion quantum computer:}
Standard, commercially available optical elements are the foundational abstraction layer of the optical system. Higher level designs can use variables or dictionaries to define optical elements with which to compile baseplates or subsystems. Where each optical element can have predefined characteristic distances and properties dynamically determining the final compiled optical layout.  
Next, some optics assemblies are stand-alone but not baseplates, such as the extended cavity diode laser (ECDL).
Next, baseplates combine individual optical elements for specific functions, such as intensity and frequency modulation with AOMs.
To be abstractable, baseplates must be modular and fit within any larger optical sub-system with minimal customization.
This means their input and output beams must conform to design rules so that all beams enter and exit along the grid of the optical table and flow in the same direction between baseplates, to simplify alignment between sequential baseplates. Next, these combinations of baseplates for various functions creates self contained subsystems at each wavelength, encompassing even the laser source. Here showing a Subsystem for cooling and detection of strontium ions including: a custom 422 nm ECDL, optical isolation, saturated absorption spectroscopy locked to Rb, then a single-pass AOM plate and a double-pass AOM plate.
Lastly, combining multiple subsystems at multiple wavelengths creates an entire apparatus. In this case, a strontium trapped ion quantum computer. Each of these subsystems utilize the same underlying modular designs but with wavelength specific differences such as optics coated at different wavelengths and different laser diodes.
}
\label{fig:full}
\end{figure*}

\begin{figure*}[]
\centering
\includegraphics[width=\textwidth]{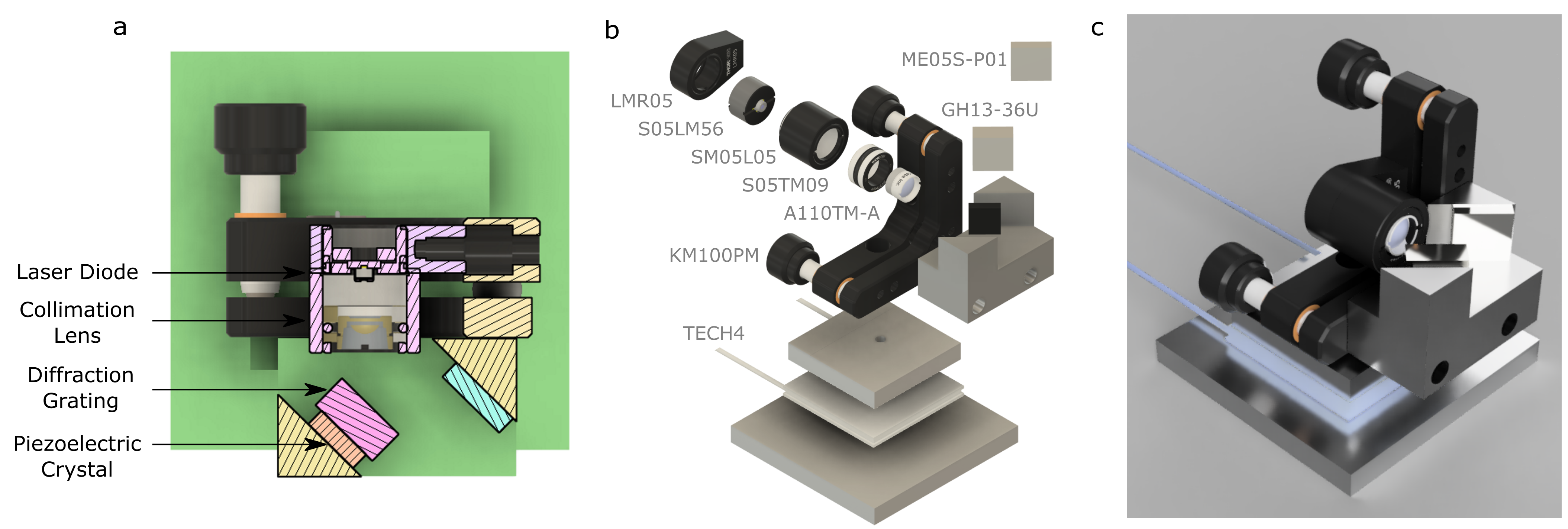}
\caption{
\textbf{Extended cavity diode laser:}
PyOpticL 422 nm extended Cavity Diode Laser (ECDL) with almost all off-the-shelf components. The only custom part is a 3D printed mount for the grating with a dynamic Littrow angle.
}
\label{fig:ECDL}
\end{figure*}

\subsection{Dynamic Designs for Layout and Routing}
PyOpticL leverages this dynamic beam path routing to create an intuitive and efficient code-to-CAD optical layout tool, allowing for dynamic and easily modified optical layouts. 
Components can be defined via coordinates on the baseplate, relative to the beam path, or relative to other components. Using these varying placement methods, it is easy to create layouts which are entirely constrained to the beam path. This method of design allows for the entire layout to dynamically adjust whenever a single element is moved or modified, and therefore automatically preserves the desired beam path and maintains connectivity. This means that many subtle adjustments can be made through the course of rapid development within the CAD environment without ruining the entire layout.

A choice to use a global set of beam directions allows mounts to be placed so that they accept one orientation of beam input and output another (e.g. `up-right' accepts a beam going 'up' and turns it `right'). These orientations are not relative to the beam path, meaning that ``up/down" and ``left/right" are defined globally with respect to the baseplate. This means that no angles are needed to script the layout. (We note that arbitrary angles are possible and certainly beneficial in particular instances like very compact layouts.)

The scripted nature of the library also allows for highly parameterized designs through the use of local variables. For instance, the type of mirror used throughout an entire layout could be easily changed simply by defining it as an argument or a variable in that layout's script. Furthermore, components can be dynamically designed to adapt to varying parameters, such as adapter plates generated on-demand for different mounting angles based on application-specific wavelengths or optical properties like grating pitch.

Code-to-CAD design therefore enables layouts to be recompiled with entirely different sets of mounts, or indeed, no mounts at all, if a full monolithic mounting is desired. To demonstrate this capability, we recompile the same saturated absorption spectroscopy (SAS) layout in four cases: 1. 1-inch mounted optics, 2. 1/2-inch mounted optics, 3. 1/2-inch unmounted optics, 4. Mini-optics , shown in Figure. \ref{fig:dynamic}. By defining a set of variables for each of these scales, a single variable can be used to switch between them.
We note that there are \textit{many} examples of small SAS setups \cite{sas_vescent, christ2019integrated, kurbis2020extended, madkhaly2021performance, christ2024additive, maurice2020miniaturized}, but they are static designs.

To further improve the interoperability of plates, we also enable global switching between metric and imperial measurement systems. When appropriate, components and mounting hardware can be configured to support both metric and imperial variants. Additionally, if aligning components to the underlying grid, a 'grid unit' is provided which is automatically set based on the measurement system. In this way, layouts should always be compatible with both systems without requiring any additional design work.

~
~

\noindent\begin{minipage}{\linewidth} 
\begin{lstlisting}[language=Python, style=myPythonStyle, basicstyle=\footnotesize, frame=single]
from PyOpticL.settings import set_measurement_system
# set global measurement system setting
set_measurement_system("metric")
\end{lstlisting}
\end{minipage}

~
~

In order to streamline the prototyping process further, PyOpticL stores part information for baseplate components and is capable of outputting a dynamically generated Bill of Materials (BOM) which is formatted so it can be easily imported to the vendor website for quickly generating quotes and purchasing.

\begin{figure*}
\centering
\includegraphics[width=\textwidth]{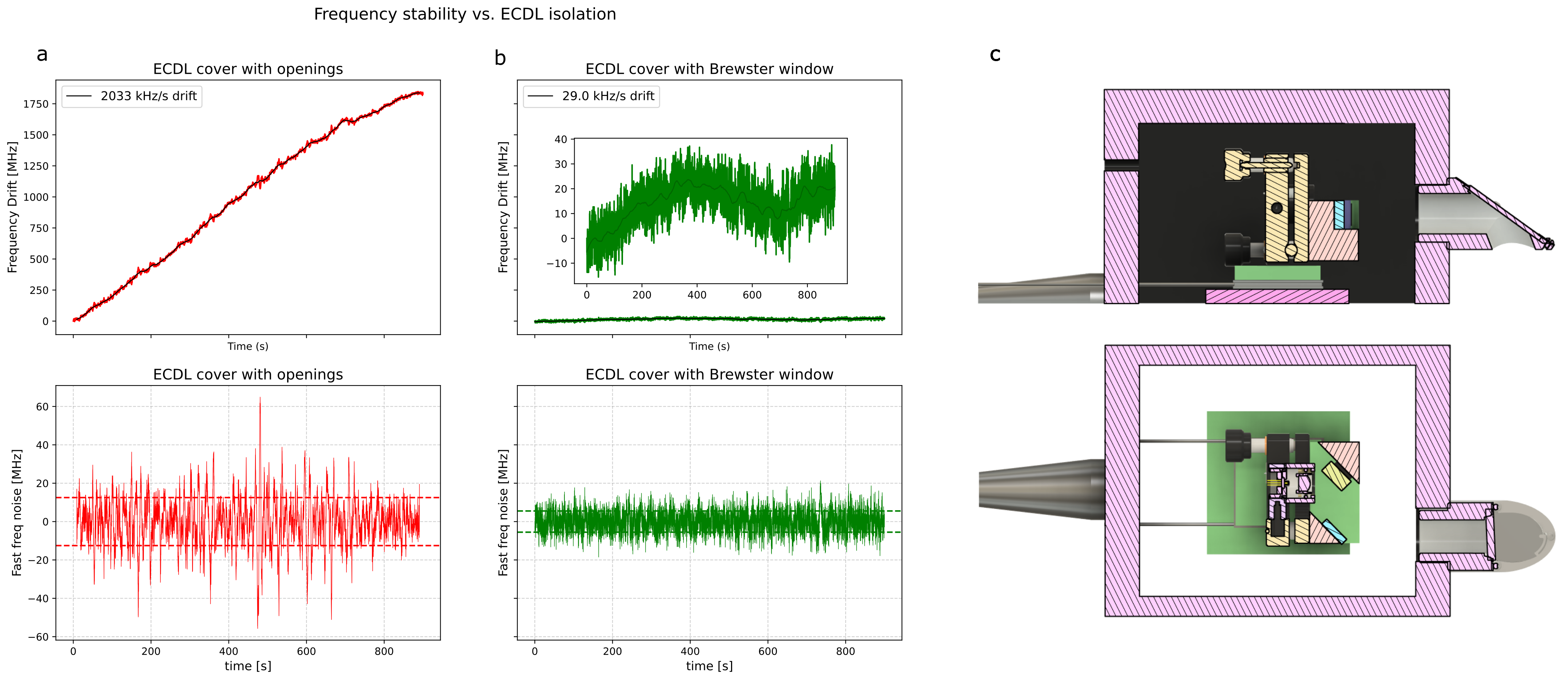}
\caption{
\textbf{ECDL passive frequency stability:}
Using a Brewster window to create a 3D printed air tight cover dramatically improves the laser stability. a) When the box for the ECDL has small openings for the laser, wires, and adjustment of the diffraction grating window we see drift of over 2MHz/s. b) When we use a Brewster window to create an air tight cover for the laser, we see this drift reduced by nearly two orders of magnitude, to 30kHz/s. Further, the fast frequency noise is reduced by half as well. c) Model of the air tight cover design that was 3D printed.
}
\label{fig:ECDL_stability}
\end{figure*}

\newpage

\subsection{Creating New Components}

One critical bottleneck in a code-to-CAD system can be the introduction of new components into the library. To make this process as accessible as possible we created a \textit{component definition} system. Component definitions are abstract classes that allow for easy declaration of component attributes in a very pythonic form. Inside the class, functions can define the shape, required drilling, optical interfaces, and any sub-components of the object. As these are standard python classes, they can also be parameterized with any number of instance arguments. For example, a generic mirror definition can be created with a number of adjustable parameters, then within a given design, an instance of this definition can be created tailored for that application. This instance can then be used repeatedly to create mirrors within the layout, all of which will have those same attributes. Further, since these definitions are not directly linked to the PyOpticL software, local libraries of components can be easily created to store design-specific components or other custom parts.

It is also very important that importing geometry for new mounts or custom parts is streamlined and precise. Otherwise, the process of aligning these parts within CAD can return to a tedious trial-and-error process, which defeats the purpose of the system entirely. 
Therefore, we have developed multiple GUI tools to allow for quick alignment, positioning, and measurement of parts during import, allowing you to store them in either the internal PyOpticL library or in a custom external library (see GitHub \cite{pyopticl} for details).
This can also be useful for importing non-optical elements like vacuum chambers, etc. to register beam delivery to an experimental apparatus.

\newpage

\section{Applications}

\subsection{Optical Subsystems for AMO and Quantum Computing Experiments}

There are a few key subsystems for optics within all atomic and molecular optical physics experiments, including quantum computing with neutral atoms and trapped ions. 

\begin{enumerate}
    \item \textbf{Laser} - Extended cavity diode lasers (ECDLs) as a source of narrow linewidth laser light.
    \item \textbf{Laser stabilization} - Locking to atomic transitions for absolute laser stabilization.
    \item \textbf{Pulse control} - Acousto-optical modulators for modulating the laser intensity and frequency. Commonly with 'single-pass' and 'double-pass' configurations. 
    \item \textbf{Beam delivery} - Finally, the light must be delivered to the target trapped ion, atom, sample, etc., with a precisely aligned lens. 
\end{enumerate}

Here we outline the development of example or template baseplates for each of these applications, which are readily reconfigurable for any wavelength and numerous AMO applications, using the PyOpticL library (Figure \ref{fig:full}). The full setup can be created by importing each baseplate as a function, matching their input and outputs manually relative to the underlying 1-inch grid of the optical table (a much easier task than aligning each optical element individually). We have found that in practice further automatic alignment is not necessary.

~
\newline
~

\noindent\begin{minipage}{\linewidth} 
\begin{lstlisting}[language=Python, style=myPythonStyle, basicstyle=\footnotesize, frame=single]
from doublepass import doublepass
from ecdl import ecdl
from rb_sas import rb_sas
from singlepass import singlepass

subsystem = Layout("Laser Cooling Subsystem")

subsystem.add(ecdl,
    position=(0, 0, 0),rotation=180)
subsystem.add(rb_sas,
    (dim(-6, "grid"), dim(-17, "grid"), 0),rotation=90)
subsystem.add(singlepass,
    (dim(-14, "grid"), dim(-6, "grid"), 0),rotation=90)
subsystem.add(doublepass,
    (dim(-20, "grid"), dim(-5, "grid"), 0),rotation=90)

if __name__ == "__main__":
    subsystem.recompute()
\end{lstlisting}
\end{minipage}

\begin{figure*}[]
\centering
\includegraphics[width=0.9\textwidth]{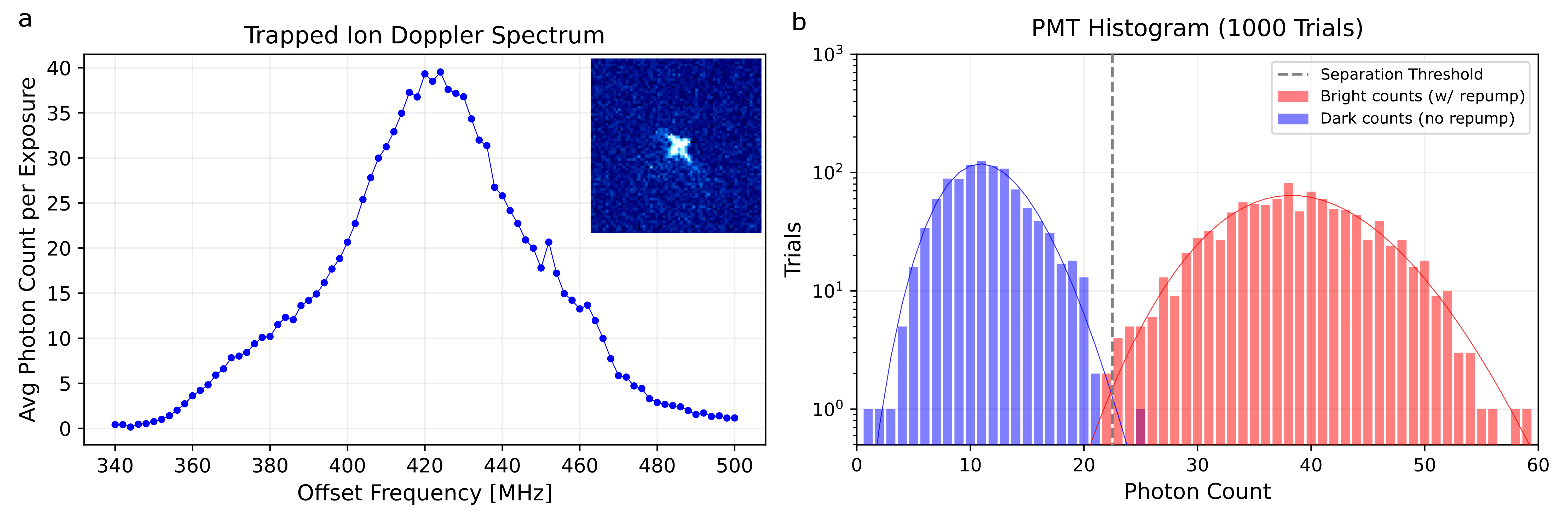}
\caption{
\textbf{Trapped ion laser cooling and detection:}
a) Doppler scan of the brightness of the trapped ion while scanning the frequency of the double-pass AOM sending light to the ion from the custom 422 nm ECDL locked to the Rb SAS (Inset: Trapped ion, cooled and detected with custom ECDL). 
b) Detection histogram of the trapped ion using the custom 422 nm ECDL shows high fidelity detection (99.8\%) in 2 ms detection time. Dark counts are measured by manually blocking the repump laser (1092 nm). 
}
\label{fig:ion}
\end{figure*}

\subsection{Dynamic Extended Cavity Diode Laser (ECDL)}

Homemade extended cavity diode lasers at near-IR wavelengths were an enabling tool for pioneering AMO laboratories \cite{arnold1998simple}, which needed many narrow linewidth lasers for various atomic transitions and operations such as laser cooling and repumping. However, these designs are all static and require the machining of multiple custom components \cite{arnold1998simple, hawthorn2001littrow,cook2012high, dutta2012mode, doret2018simple, chang2023low, schkolnik2019extended, daffurn2021simple, kurbis2020extended}.

Here we focus on designing an extremely simple ECDL with only a single custom part, a dynamic 3D printed mount for the diffraction grating, generated in PyOpticL. All other parts are off-the-shelf with no modification or machining at all. The 3D printed grating mount attaches directly to the ubiquitous KM100PM Thorlabs mount without modification. As an example, we demonstrate laser operations at 421.6 nm, which is useful for strontium trapped ions and also neutral rubidium for Rydberg applications. 

We mount a 418 nm laser diode (TopGaN, no AR coating) within a 1/2 inch lens tube (SM05L05), threaded to the front of the LMR05, which is directly mounted to a KM100PM (Figure \ref{fig:ECDL}). 
The KM100PM is mounted onto a small aluminum block, which is hand cut from 1/4 inch stock with a single drilled and 8-32 threaded hole.
A TEC (TECH4) is then sandwiched with another aluminum block and controlled by a TED200C (see SI for more details).

The feedback for the extended cavity diode laser is provided by a diffraction grating (GH13-36U, 7\% at 421nm) mounted on a parameterized custom 3D printed mount generated by PyOpticL. The Littrow angle of the grating is set by the wavelength and the diffraction grating pitch, allowing the mount design to be dynamically reconfigured for arbitrary wavelengths.
We have simplified this design to minimize the amount of machining required. This mount can be milled from aluminum with a single set angle of milling and drilling two holes (with counter-sink). However, we found no need to fabricate an aluminum version as the 3D printed mount shows remarkable stability and was used for all experiments. The material and parameters used to print this mount were the same as all other small adapters (see Rapid Prototyping with 3D Printing).


We tuned the laser diode free-running wavelength of 418.9 nm up to 421.6 nm (-4.6 THz) by heating it to 50 C.
Thermally isolating the laser within a 3D printed box enabled stability within 100MHz for hours measured by a MOGLabs Fizeau wavemeter and verified with saturated absorption spectroscopy.

To further isolate and stabilize the laser we created a totally enclosed 3D printed box with a port for a Brewster window for light to exit. This dramatically improved the laser stability over the `open-to-air' box (Figure \ref{fig:ECDL_stability}), reducing drift from 2MHz/s to 30kHz/s and reducing fast frequency noise.

\begin{figure*}[]
\centering
\includegraphics[width=\textwidth]{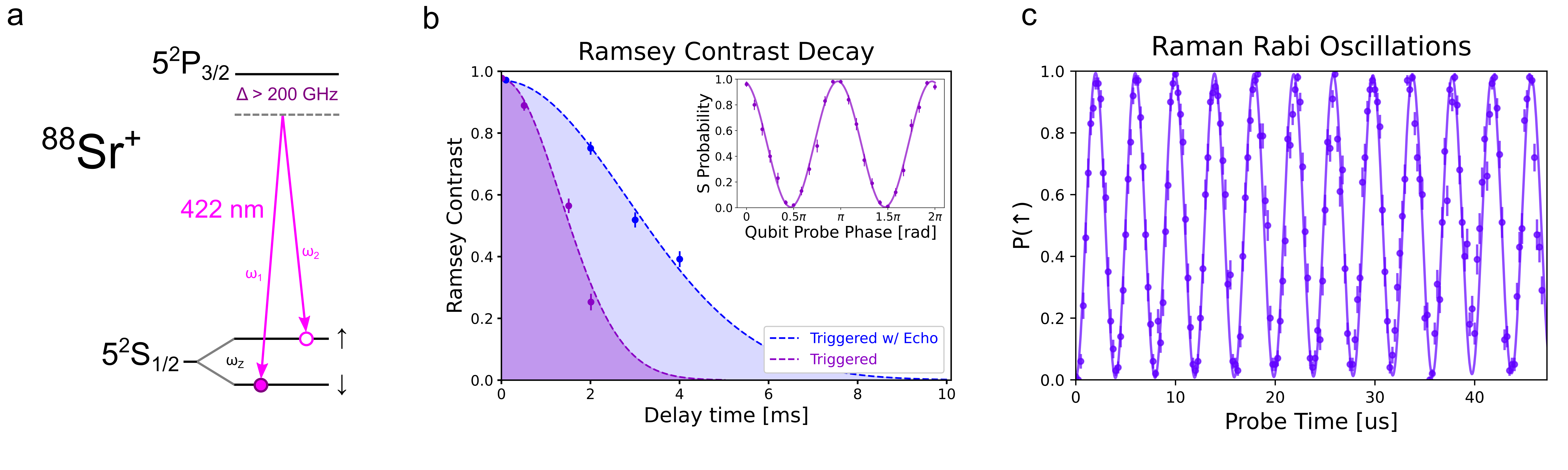}
\caption{
\textbf{Zeeman qubit operations with Raman laser:}
a) Sublevels of the ground state of \Sr are Zeeman split by $\omega_Z$ from an applied 6 Gauss magnetic field. The Raman beams are detuned by $\Delta > 200$ GHz from the excited state, and their relative frequency is set to resonance with the splitting of the ground state. b) We measure a coherence time of 1.88 ms by performing Ramsey interferometry triggered to AC mains, with increasing delays for free evolution. We are able to extend this time to 4.00 ms by performing a spin echo with the Raman laser c) Raman Rabi oscillations retain $>99\%$ spin population per inversion beyond 20$\pi$ rotations.
}
\label{fig:qubit}
\end{figure*}

\subsection{Laser Stabilization with Saturated Absorption Spectroscopy}

To demonstrate stable single-mode operation of the ECDL we use it to observe saturated absorption spectroscopy (SAS). SAS is commonly used to lock and stabilize lasers using an atomic transition. We note that \textit{many} other examples of small SAS setups exist \cite{knappe2007microfabricated, kurbis2020extended, madkhaly2021performance, christ2024additive} and are commercially available \cite{sas_thorlabs,sas_vescent}.

Here we create saturated absorption spectroscopy (SAS) layouts using PyOpticL (Fig.\ref{fig:dynamic}), and build the 1/2 inch mounted optics plate (see SI for more details).
Tuning the laser to the resonance of rubidium enabled capture of Doppler-free peaks (SI Figure \ref{fig:SAS}a). 

Next, we stabilize the ECDL frequency by locking to the $\text{5s}^2\text{S}_{1/2} \rightarrow 6\text{p}^2\text{P}_{1/2}$ transitions $^{85}\text{Rb}$  (F = 2 $\rightarrow$ 2,3) \cite{akerman2012trapped,madej1998rb} using the open-source software Linien \cite{wiegand2022linien} and a Red Pitaya. Locking feedback was applied through the voltage to the PZT, and modulation was added via the Thorlabs current controller.
We observed that the lock was stable for hours at a time over the course of weeks.


\subsection{Laser Cooling and Detection}

With the ECDL locked to the rubidium (F = 2 $\rightarrow$ 2,3) transition by the SAS baseplate, we then used the double-pass AOM baseplate to control the laser intensity and frequency, spanning the gap to resonance with the strontium ion ($5\text{S}_{ 1/2} \rightarrow 5\text{P}_{ 1/2}$ transition) for Doppler cooling and detection \cite{jung2017all,akerman2012trapped,madej1998rb}. 

We used this complete PyOpticL optical subsystem to laser-cool trapped \Sr ions (Figure \ref{fig:ion}a, inset) for hours at a time over the course of weeks. 
To verify precise frequency control via the double-pass AOM baseplate we also measure the brightness of the ion while scanning across resonance, measuring the Doppler curve of the trapped ion (Figure \ref{fig:ion}a). This also verifies that the SAS lock baseplate stabilized the laser to within a few MHz. Finally, we verified that the system can perform high-fidelity state detection (Figure \ref{fig:ion}b), measuring 99.8\% fidelity with a detection time of 2 ms. The dark counts are measured by manually blocking a repump laser.






\subsection{Raman Laser for Zeeman Qubit Gate Operations}

Using the same ECDL but a different 422 nm diode (Nichia) we built a Raman laser for Zeeman qubit gate operations between the spin sublevels of the ground state (Fig. \ref{fig:qubit}a). We used the single-pass AOM baseplates to create two tones ($\omega_1,\omega_2$) with independent intensity and frequency control and combined them into the same optical fiber for co-propagating delivery to the ion.
To create a Zeeman qubit \cite{ruster2016long}, we then set the difference of the two tones equal to the frequency of the Zeeman splitting between the sublevels ($\delta = \omega_Z - (\omega_1 - \omega_2)= 0$) and pulsed the Raman beams to excite spin transitions of the ground state (Fig. \ref{fig:qubit}b).
To test qubit operations, we tuned the Raman laser 200~GHz from resonance (to reduce the error from spontaneous emission \cite{ozeri2007errors} to below 0.1\%) and were able to measure Raman Rabi oscillations (Fig. \ref{fig:qubit}c), equivalent to a single-qubit gate fidelity of greater than $99\%$. We estimate the gate fidelity by fitting the Rabi oscillations to a decaying sinusoid with a linear decay envelope (valid for the short times relative to the $\tau_2$ coherence time). From this we extract a SPAM fidelity of $98.5\% \pm 1.0\%$ \cite{chauhan_trapped_2024}, and a fidelity per inversion of $99.95\% \pm 0.03\%$.
We also measure magnetic field noise limited coherence times of $1.88 \pm 0.05$ ms when triggering experiments to the AC mains, which can be extended to $4.0 \pm 0.1$ ms with a spin-echo pulse.(Fig. \ref{fig:qubit}b). In the future, adding Mu-metal shielding could extend it beyond a second \cite{ruster2016long}.


\begin{figure*}[]
\centering
\includegraphics[width=\textwidth]{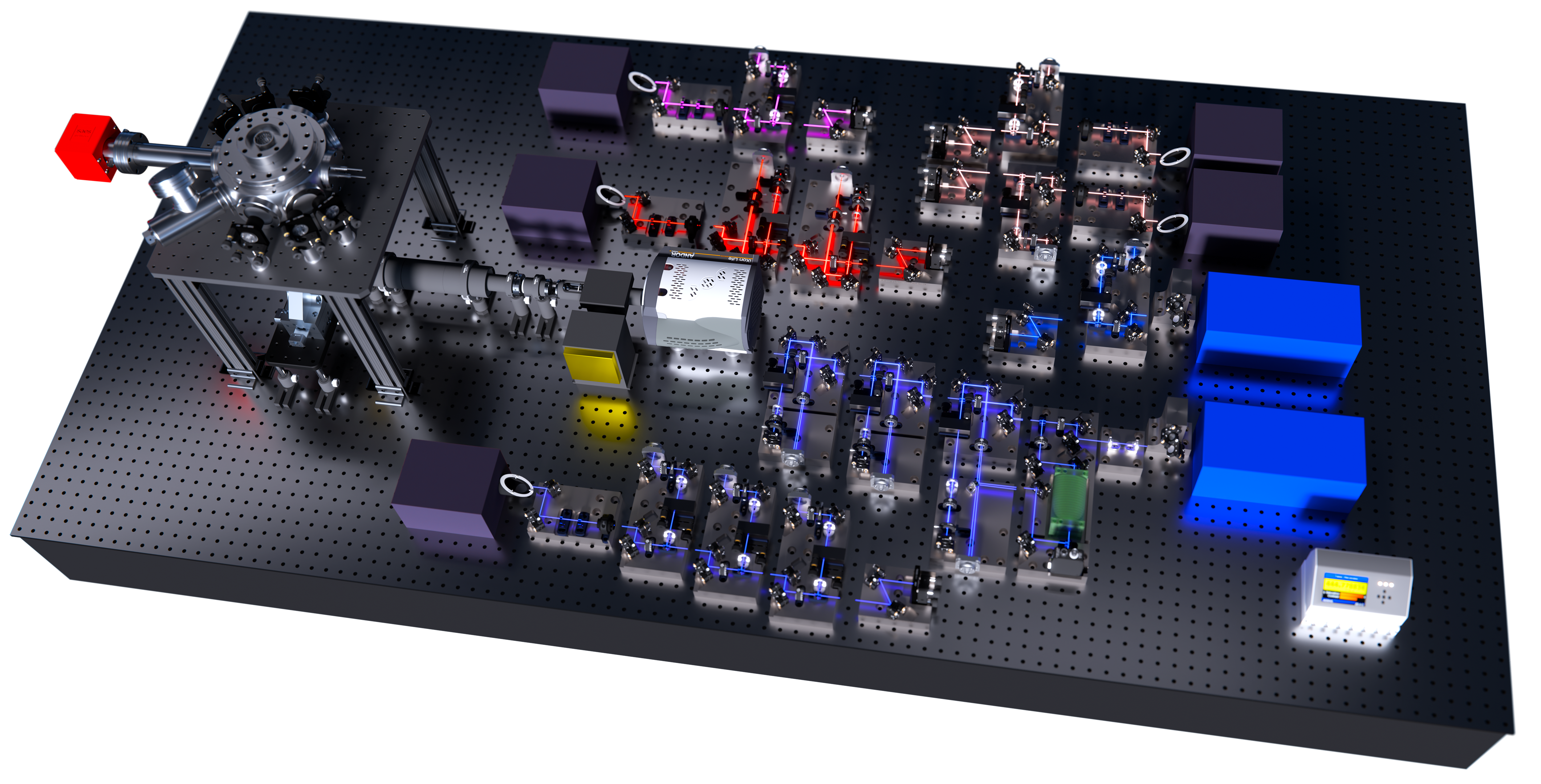}
\caption{
\textbf{Full PyOpticL Apparatus with subsystems comprised of identical modular baseplates:}
Strontium trapped ion system with modular optical system created with PyOpticL. 
}
\label{fig:fullsystem}
\end{figure*}

\subsection{Dynamic Laser System Demonstration}

Altogether, we have demonstrated a fully dynamic AMO apparatus (Figure \ref{fig:full}) with modular optical subsystems that are dynamically reconfigurable for innumerable AMO applications, including quantum computing with trapped ions and neutral atoms. 
Each optical subsystem can be recompiled for any wavelength and, therefore, any atomic species. 
This means that nearly identical subsystems to those we have demonstrated could be used for a neutral-atom AMO apparatus. 
For example, a Bose-Einstein condensate apparatus \cite{olson2014tunable, li2019spin} could use the same cooling and detection subsystem layouts that we have demonstrated but with optics for 780 nm (for Rubidium atoms), and a duplicate subsystem could be used for repumping.
Even more directly, the exact same laser cooling subsystem that we have demonstrated could be used in a neutral-atom quantum computer based on Rb \cite{bluvstein2024logical} as they also require lasers at 421 nm and lock to Rb for stabilization.

This modularization and standardization of optical circuits begins to enable design abstraction methods similar to those used in digital logic systems. Where details of the underlying optical baseplate are abstracted away, with only their function in the optical system being important for design. This paradigm is critical not only for rapid and sustainable development, but also for scaling. VLSI would not scale without abstraction or without modularization and standardization of the underlying subsystems.

\begin{figure*}[]
\centering
\includegraphics[width=\textwidth]{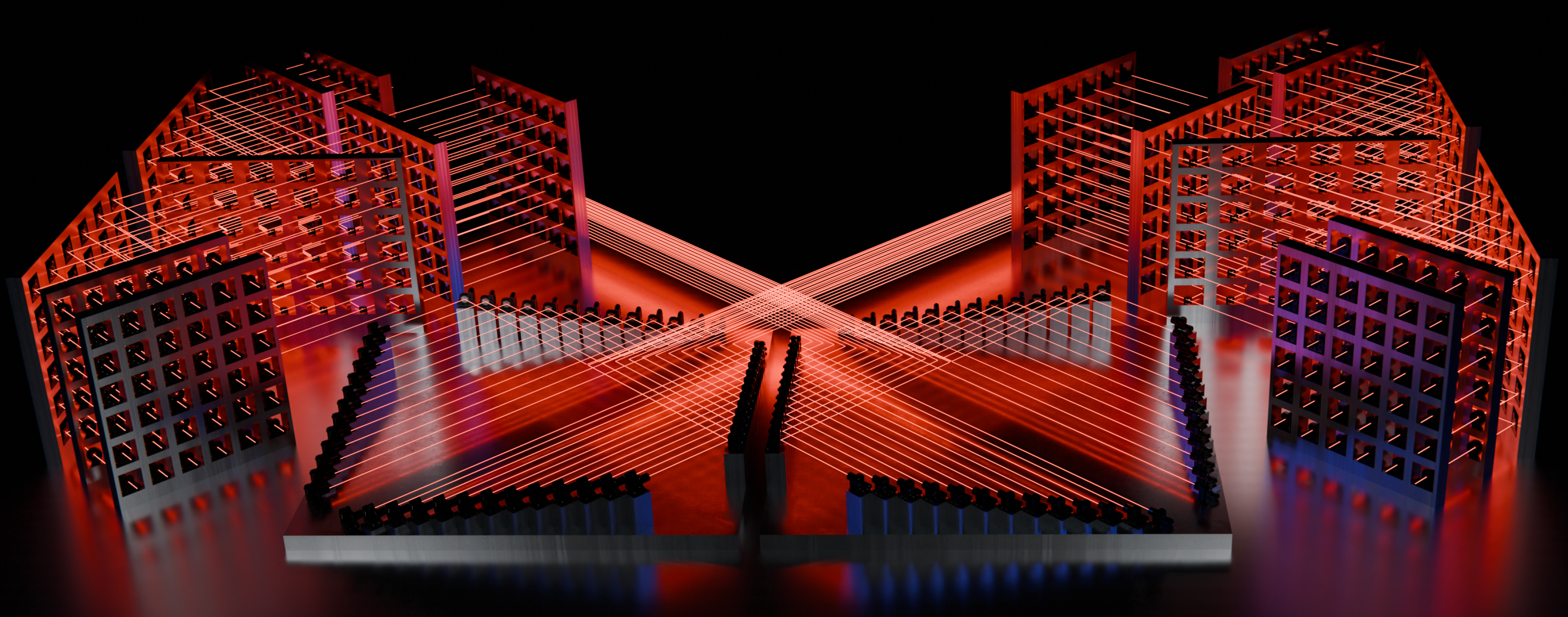}
\caption{
\textbf{Scalable optical layout:}
To showcase the scalable nature of the code-to-CAD design paradigm we roughly re-create the optical layout used in a recent optical interferometry experiment \cite{zhong2021phase}; using for-loops to layout the individual optical elements in grid mounts tracking the laser beam for each path. We then render the CAD model in Blender. 
}
\label{fig:redstone}
\end{figure*}

\section{Rapid Prototyping with 3D Printing }

Over the course of this project prototypes of optical baseplates were printed using an Elegoo Saturn 3D resin printer, with engineering resin (or high-temperature resin for the heated vapor cell holder). Due to the speed of fabrication allowed by 3D printing, multiple prototype designs can be quickly fabricated and tested to verify placement of all optics before machining a metal version. The modular nature of the baseplates with respect to each other makes them very easy to swap and re-organize as needed without realigning their constituent optical elements. (see SI section "3D printing procedures and best practices" for more details)

All baseplates are designed to be easily CNC milled out of aluminum to maintain the stability of the underlying commercial optical mounts and maximize the stability of the full layout.
This means that all designs use a subtractive drilling operation within PyOpticL so that baseplates can be easily machined from 1 inch thick stock aluminum.
PyOpticL automatically generates technical CAD drawings of the designs, including annotations noting threaded holes etc. to streamline ordering and machining.

We observe that optics mounted on 3D printed prototype plates exhibit a settling period of approximately one week. Daily optical realignment is required during this period to maintain power efficiency. After the initial settling, the power drop stabilizes and only minor weekly realignment is needed to prevent significant beamline drift. 
In contrast, plates that are CNC milled from aluminum have not required regular realignment during our use. For this reason, although 3D printed plates can be reliably maintained for experimental work over periods of several months, it is advisable to transition to metal versions once a final design has been reached.



For mounting individual optics like PBS and rotation mounts, small 3D printed fixed adapters acting as sub-mounts have been sufficient without aluminum machining, as their small size limits deformation. 

To further quantify this stability we measured the optical power fiber-coupled through a Doppler cooling subsystem beam path composed of 3D printed adapter mounts and a mix of 3D printed and CNC-machined baseplates. We observe that over a 12-hour period there is a 2\% drop in optical power (see Fig. \ref{fig:PowerStability} in the SI), which is consistent with our experience prototyping with PyOpticL over the past two years.

It is interesting to note that for some applications, 3D printing a `mount free' baseplate without any commercial adjustable mounts could be sufficient and much more compact (Fig. \ref{fig:dynamic}c), dramatically lowering the cost for these optical systems and making them more accessible for broader educational applications. 

While we anticipate the future development of 3D printed adjustable mounts within the PyOpticL library to replace the commercial adjustable mounts, we are cautious that the compromise in performance with entirely 3D printed adjustable mounts for most applications would be unjustifiable. 

\section{Scalable Optical Design}
Script-based optical design enables new paradigms of scalable optical layout through inheritance of the scalable nature of code. To showcase this, we used PyOpticL to roughly recreate the layout (Figure \ref{fig:redstone}) from an optical interferometry experiment \cite{zhong2021phase} designed to demonstrate quantum supremacy through Boson Sampling. 

The goal of this circuit is to facilitate the simultaneous interferometry of a 144 mode photonic circuit. To do this, an array of input lasers were routed into the photonic circuit, the output of which was then split and measured using a layout of custom vertical arrays ( 6 x 6 ) of optical elements. To model this, the surface and grid mounts were created as custom components, with the latter being designed such as to be dynamically generated with arbitrary array dimensions. From there, the entire layout could be designed using simple for loops to place the large number of optics required. Hence, the entire design becomes dynamically scalable giving the option of vastly increasing the number of photonic modes through a single parameter change. While the fabrication and implementation of such large photonic circuits may not be feasible, we hope that this example from a famous quantum optics experiment inspires more creative applications of PyOpticL for sophisticated optical layouts which can take advantage its scalable design philosophy.

We also note that this scaling is critical for the broader adoption of micro-optics, which are more practically capable of creating such large optical setups. 




\section{Discussion}

We hope that PyOpticL becomes a hub for open-source hardware development of optical layouts and optical systems, as more baseplates and layouts are created for new applications. We also hope that this demonstration of a full laser system comprised of multiple modular sub-systems including: laser sources, spectroscopy and stabilization, and modulation baseplates for laser cooling, detection, and qubit gates with trapped ions provides a solid foundation that is already applicable to a broad array of AMO experiments. 

Our ECDL design is one fifth of the cost of commercial lasers, has the same linewidth (55 kHz), and can be assembled with off-the-shelf components (except for a single 3D printed or machined part) in minutes after next-day shipping from Thorlabs, dramatically lowering the barrier to entry into AMO and quantum computing for education and research.
Total costs for our ECDL are $\$6,806$, with almost half of that cost being the relatively expensive 420 nm laser diode itself ($\$$3k) and most of the rest being the optical isolator and commercial current, temperature, and PZT controllers, which could be swapped for homemade designs \cite{doret2018simple} if reducing cost is paramount. The Thorlabs parts that comprise the ECDL itself cost \$596 (see SI for more details), which is only a tenth of the total cost. 
Demonstration of locking to an atomic cell, qubit operations, and high fidelity qubit gates shows that these lasers can be used in precision AMO experiments. 

In the future, we also hope to continue to pursue more micro-optic layouts utilizing novel 3D printing techniques with materials that are stable enough for long-term operation. 
We are also optimistic that more advanced additive manufacturing could enable the integration of the UHV chamber \cite{cooper2019additively} directly into the PyOpticL library, which in turn could enable creative and unconstrained layouts for optical delivery. 

\section{Community}

We also envision PyOpticL contributing to a broader, fully open-source quantum computing ecosystem together with projects such as M-Labs ARTIQ and Sinara \cite{kasprowicz2020artiq}. This will help increase the accessibility and reproducibility of quantum research. In its first year, the PyOpticL community has grown to include 11 experimental groups across 9 institutions worldwide in both academia and industry. Several groups are already extending the library beyond our initial trapped-ion demonstrations to support new neutral-atom platforms.

We are particularly encouraged to see groups collaborating with us and with each other as they develop their systems. Through this process, we hope the library can help converge best practices for optical design and integration. As adoption grows, we are optimistic that PyOpticL will accelerate the pace of research in the community by simplifying experimental setup, enabling modular and reconfigurable architectures, and fostering a new model of collaborative optical system design.

\section{Author Contributions}
R.J.N. conceived of the work.
J.M., J.O., and R.J.N. wrote the PyOpticL library.
N.H. and Z.W. performed the SAS and trapped ion experiments with assistance from C.C. and R.J.N. 
The Raman qubit experiments were performed by C.C., Z.W., and N.H. with assistance from R.J.N. 
All authors contributed to developing optical baseplates using the library. 
All authors discussed the results and contributed to the writing of the paper. 
R.J.N. supervised the research.

\section{Acknowledgments}
This material is based upon work supported by the National Science Foundation under Grant No. (2338369).
The authors gratefully acknowledge Professor Isaac Chuang for inspiring this work. He and his team at the MIT Quanta Lab developed an optical layout library within OpenSCAD called CAD for Precision Optics (C4PO) which inspired this work.

\bibliographystyle{unsrtnat}
\bibliography{references} 

@misc{pyopticl,
	title = {{PyOpticL}},
	howpublished = "\url{https://github.com/UMassIonTrappers/PyOpticL}",
}

@misc{gdsfactory,
	title = {gdsfactory},
	howpublished = "\url{https://gdsfactory.github.io/gdsfactory/index.html}",
}

@misc{freecad,
	title = {{FreeCAD}: {Your} own {3D} parametric modeler},
        shorttitle = {{FreeCAD}},
	howpublished = "\url{https://www.freecad.org/}",
}

@misc{duke,
	title = {Duke {University} - {Compact} {Ion} {Trap} {System}},
        journal = {EURIQA},
	howpublished = "\url{https://euriqa.pratt.duke.edu/research/compact-ion-trap-system}",
}

@misc{JayichUCSBdoublepass,
	title = {{Jayich Lab UCSB} double-pass-breadboard},
	howpublished = "\url{https://github.com/Jayich-Lab/double-pass-breadboard}",
}

@misc{aosense,
	title = {{AOSense} rack laser system},
	howpublished = "\url{https://aosense.com/products/lasers/rack-laser-system/}",
}

@misc{AQTrowan,
	title = {{AQT Rowan} rack-laser-system},
	howpublished = "\url{https://www.aqt.eu/rowan/}",
}

@article{chen2022stable,
  title={Stable turnkey laser system for a yb/ba trapped-ion quantum computer},
  author={Chen, Tianyi and Kim, Junki and Kuzyk, Mark and Whitlow, Jacob and Phiri, Samuel and Bondurant, Brad and Riesebos, Leon and Brown, Kenneth R and Kim, Jungsang},
  journal={IEEE Transactions on Quantum Engineering},
  volume={3},
  pages={1--8},
  year={2022},
  publisher={IEEE},
  doi={https://doi.org/10.1109/TQE.2022.3195428}
}

@article{spivey2021high,
  title={High-stability cryogenic system for quantum computing with compact packaged ion traps},
  author={Spivey, Robert Fulton and Inlek, Ismail Volkan and Jia, Zhubing and Crain, Stephen and Sun, Ke and Kim, Junki and Vrijsen, Geert and Fang, Chao and Fitzgerald, Colin and Kross, Steffen and others},
  journal={IEEE Transactions on Quantum Engineering},
  volume={3},
  pages={1--11},
  year={2021},
  publisher={IEEE},
  doi={https://doi.org/10.1109/TQE.2021.3125926}
}

@article{pogorelov2021compact,
  title={Compact ion-trap quantum computing demonstrator},
  author={Pogorelov, Ivan and Feldker, Thomas and Marciniak, Ch D and Postler, Lukas and Jacob, Georg and Krieglsteiner, Oliver and Podlesnic, Verena and Meth, Michael and Negnevitsky, Vlad and Stadler, Martin and others},
  journal={PRX Quantum},
  volume={2},
  number={2},
  pages={020343},
  year={2021},
  publisher={APS},
  doi={https://doi.org/10.1103/PRXQuantum.2.020343}
}

@article{kulkarni2020ultrastable,
  title={Ultrastable optical components using adjustable commercial mirror mounts anchored in a ULE spacer},
  author={Kulkarni, Soham and Umi{\'n}ska, Ada and Gleason, Joseph and Barke, Simon and Ferguson, Reid and Sanju{\'a}n, Jose and Fulda, Paul and Mueller, Guido},
  journal={Applied optics},
  volume={59},
  number={23},
  pages={6999--7003},
  year={2020},
  publisher={Optica Publishing Group},
  doi={https://doi.org/10.1364/AO.395831}
}

@article{zhang2022design,
  title={Design of a highly reliable and low-cost optical bench for laser cooling},
  author={Zhang, Zhen and Xiang, Jingfeng and Meng, Yiming and Ren, Wei and Deng, Siminda and L{\"u}, Desheng},
  journal={Optical Fiber Technology},
  volume={72},
  pages={102974},
  year={2022},
  publisher={Elsevier},
  doi={https://doi.org/10.1016/j.yofte.2022.102974}
}

@inproceedings{kasprowicz2020artiq,
  title={Artiq and sinara: Open software and hardware stacks for quantum physics},
  author={Kasprowicz, Grzegorz and Kulik, Pawe{\l} and Gaska, Michal and Przywozki, Tomasz and Pozniak, Krzysztof and Jarosinski, Jakub and Britton, Joseph W and Harty, Thomas and Balance, Chris and Zhang, Weida and others},
  booktitle={Quantum 2.0},
  pages={QTu8B--14},
  year={2020},
  organization={Optica Publishing Group},
  doi={https://doi.org/10.1364/QUANTUM.2020.QTu8B.14}
}

@article{machado2019parametric,
  title={Parametric CAD modeling for open source scientific hardware: Comparing OpenSCAD and FreeCAD Python scripts},
  author={Machado, Felipe and Malpica, Norberto and Borromeo, Susana},
  journal={Plos one},
  volume={14},
  number={12},
  pages={e0225795},
  year={2019},
  publisher={Public Library of Science San Francisco, CA USA},
  doi={https://doi.org/10.1371/journal.pone.0225795}
}

@article{riegel2016freecad,
  title={FreeCAD},
  author={Riegel, Juergen and Mayer, Werner and van Havre, Yorik},
  journal={Freecadspec2002. pdf},
  year={2016}
}

@article{arnold1998simple,
  title={A simple extended-cavity diode laser},
  author={Arnold, AS and Wilson, JS and Boshier, MG},
  journal={Review of Scientific Instruments},
  volume={69},
  number={3},
  pages={1236--1239},
  year={1998},
  publisher={American Institute of Physics},
  doi={https://doi.org/10.1063/1.1148756}
}

@article{hawthorn2001littrow,
  title={Littrow configuration tunable external cavity diode laser with fixed direction output beam},
  author={Hawthorn, CJ and Weber, KP and Scholten, Robert E},
  journal={Review of scientific instruments},
  volume={72},
  number={12},
  pages={4477--4479},
  year={2001},
  publisher={American Institute of Physics},
  doi={https://doi.org/10.1063/1.1419217}
}

@article{cook2012high,
  title={High passive-stability diode-laser design for use in atomic-physics experiments},
  author={Cook, Eryn C and Martin, Paul J and Brown-Heft, Tobias L and Garman, Jeffrey C and Steck, Daniel A},
  journal={Review of Scientific Instruments},
  volume={83},
  number={4},
  year={2012},
  publisher={AIP Publishing},
  doi={https://doi.org/10.1063/1.3698003}
}

@article{dutta2012mode,
  title={Mode-hop-free tuning over 135 GHz of external cavity diode lasers without antireflection coating},
  author={Dutta, Sourav and Elliott, DS and Chen, Yong P},
  journal={Applied Physics B},
  volume={106},
  pages={629--633},
  year={2012},
  publisher={Springer},
  doi={https://doi.org/10.1007/s00340-011-4841-4}
}

@article{doret2018simple,
  title={Simple, low-noise piezo driver with feed-forward for broad tuning of external cavity diode lasers},
  author={Doret, S Charles},
  journal={Review of Scientific Instruments},
  volume={89},
  number={2},
  year={2018},
  publisher={AIP Publishing},
  doi={https://doi.org/10.1063/1.5009643}
}

@article{chang2023low,
  title={Low-drift-rate external cavity diode laser},
  author={Chang, Eddie H and Rivera, Jared and Bostwick, Brian and Schneider, Christian and Yu, Peter and Hudson, Eric R and Hunter Collaboration and others},
  journal={Review of Scientific Instruments},
  volume={94},
  number={4},
  year={2023},
  publisher={AIP Publishing},
  doi={https://doi.org/10.1063/5.0079210}
}

@article{zhong2021phase,
  title={Phase-programmable gaussian boson sampling using stimulated squeezed light},
  author={Zhong, Han-Sen and Deng, Yu-Hao and Qin, Jian and Wang, Hui and Chen, Ming-Cheng and Peng, Li-Chao and Luo, Yi-Han and Wu, Dian and Gong, Si-Qiu and Su, Hao and others},
  journal={Physical review letters},
  volume={127},
  number={18},
  pages={180502},
  year={2021},
  publisher={APS},
  doi={https://doi.org/10.1103/PhysRevLett.127.180502}
}

@article{schkolnik2019extended,
  title={An extended-cavity diode laser at 497 nm for laser cooling and trapping of neutral strontium},
  author={Schkolnik, Vladimir and Fartmann, Oliver and Krutzik, Markus},
  journal={Laser Physics},
  volume={29},
  number={3},
  pages={035802},
  year={2019},
  publisher={IOP Publishing},
  doi={10.1088/1555-6611/aaffc8}
}

@article{daffurn2021simple,
  title={A simple, powerful diode laser system for atomic physics},
  author={Daffurn, Andrew and Offer, Rachel F and Arnold, Aidan S},
  journal={Applied Optics},
  volume={60},
  number={20},
  pages={5832--5836},
  year={2021},
  publisher={Optica Publishing Group},
  doi={https://doi.org/10.1364/AO.426844}
}

@article{kurbis2020extended,
  title={Extended cavity diode laser master-oscillator-power-amplifier for operation of an iodine frequency reference on a sounding rocket},
  author={K{\"u}rbis, Ch and Bawamia, Ahmad and Krueger, Mandy and Smol, R and Peters, A and Wicht, A and Tr{\"a}nkle, G},
  journal={Applied Optics},
  volume={59},
  number={2},
  pages={253--262},
  year={2020},
  publisher={Optica Publishing Group},
  doi={https://doi.org/10.1364/AO.379955}
}

@article{cooper2019additively,
    title = {Additively manufactured ultra-high vacuum chamber for portable quantum technologies},
    journal = {Additive Manufacturing},
    volume = {40},
    pages = {101898},
    year = {2021},
    issn = {2214-8604},
    doi = {https://doi.org/10.1016/j.addma.2021.101898},
    url = {https://www.sciencedirect.com/science/article/pii/S2214860421000634},
    author = {N. Cooper and L.A. Coles and S. Everton and I. Maskery and R.P. Campion and S. Madkhaly and C. Morley and J. O’Shea and W. Evans and R. Saint and P. Krüger and F. Oručević and C. Tuck and R.D. Wildman and T.M. Fromhold and L. Hackermüller},
    doi={https://doi.org/10.1016/j.addma.2021.101898}
}

@article{knappe2007microfabricated,
  title={Microfabricated saturated absorption laser spectrometer},
  author={Knappe, Svenja A and Robinson, Hugh G and Hollberg, Leo},
  journal={Optics express},
  volume={15},
  number={10},
  pages={6293--6299},
  year={2007},
  publisher={Optica Publishing Group},
  doi={https://doi.org/10.1364/OE.15.006293}
}

@article{maurice2020miniaturized,
  title={Miniaturized optical frequency reference for next-generation portable optical clocks},
  author={Maurice, Vincent and Newman, Zachary L and Dickerson, Susannah and Rivers, Morgan and Hsiao, James and Greene, Phillip and Mescher, Mark and Kitching, John and Hummon, Matthew T and Johnson, Cort},
  journal={Optics Express},
  volume={28},
  number={17},
  pages={24708--24720},
  year={2020},
  publisher={Optica Publishing Group},
  doi={https://doi.org/10.1364/OE.396296}
}

@article{wiegand2022linien,
  title={Linien: A versatile, user-friendly, open-source FPGA-based tool for frequency stabilization and spectroscopy parameter optimization},
  author={Wiegand, Benjamin and Leykauf, Bastian and J{\"o}rdens, Robert and Krutzik, Markus},
  journal={Review of Scientific Instruments},
  volume={93},
  number={6},
  year={2022},
  publisher={AIP Publishing},
  doi={https://doi.org/10.1063/5.0090384}
}

@article{strangfeld2021prototype,
  title={Prototype of a compact rubidium-based optical frequency reference for operation on nanosatellites},
  author={Strangfeld, Aaron and Kanthak, Simon and Schiemangk, Max and Wiegand, Benjamin and Wicht, Andreas and Ling, Alexander and Krutzik, Markus},
  journal={JOSA B},
  volume={38},
  number={6},
  pages={1885--1891},
  year={2021},
  publisher={Optica Publishing Group},
  doi={https://doi.org/10.1364/JOSAB.420875}
}

@misc{sas_thorlabs,
	title = {Thorlabs Saturated Absorption Spectroscopy Systems},
	howpublished = "\url{https://www.thorlabs.com/newgrouppage9.cfm?objectgroup_id=5616}",
}

@article{christ2024additive,
    title={Additively Manufactured Ceramics for Compact Quantum Technologies},
    author={Christ, Marc and Zimmermann, Conrad and Neinert, Sascha and Leykauf, Bastian and Döringshoff, Klaus and Krutzik, Markus},
    journal={Advanced Quantum Technologies},
    volume={7},
    number={12},
    pages={2400076},
    year={2024},
    url={https://advanced.onlinelibrary.wiley.com/doi/abs/10.1002/qute.202400076},
    doi={https://doi.org/10.1002/qute.202400076},
}

@article{madkhaly2021performance,
  title={Performance-optimized components for quantum technologies via additive manufacturing},
  author={Madkhaly, SH and Coles, LA and Morley, C and Colquhoun, CD and Fromhold, TM and Cooper, N and Hackerm{\"u}ller, L},
  journal={PRX Quantum},
  volume={2},
  number={3},
  pages={030326},
  year={2021},
  publisher={APS},
  doi={https://doi.org/10.1103/PRXQuantum.2.030326}
}

@article{christ2019integrated,
  title={Integrated atomic quantum technologies in demanding environments: development and qualification of miniaturized optical setups and integration technologies for UHV and space operation},
  author={Christ, Marc and Kassner, Alexander and Smol, Robert and Bawamia, Ahmad and Heine, Hendrik and Herr, Waldemar and Peters, Achim and Wurz, Marc Christopher and Rasel, Ernst Maria and Wicht, Andreas and others},
  journal={CEAS Space Journal},
  volume={11},
  number={4},
  pages={561--566},
  year={2019},
  publisher={Springer},
  doi={https://doi.org/10.1007/s12567-019-00252-0}
}

@misc{sas_vescent,
	title = {Vescent Saturated Absorption Spectroscopy Systems},
	howpublished = "\url{https://www.vescent.com/manuals/doku.php?id=d2:spectroscopy_module_210}",
}

@article{jung2017all,
  title={All-diode-laser cooling of Sr+ isotope ions for analytical applications},
  author={Jung, Kyunghun and Yamamoto, Kazuhiro and Yamamoto, Yuta and Miyabe, Masabumi and Wakaida, Ikuo and Hasegawa, Shuichi},
  journal={Japanese Journal of Applied Physics},
  volume={56},
  number={6},
  pages={062401},
  year={2017},
  publisher={IOP Publishing},
  doi={https://doi.org/10.7567/JJAP.56.062401}
}

@phdthesis{akerman2012trapped,
  title={Trapped ions and free photons},
  author={Akerman, Nitzan},
  year={2012},
  school={The Weizmann Institute of Science (Israel)}
}

@article{madej1998rb,
  title={Rb atomic absorption line reference for single Sr+ laser cooling systems},
  author={Madej, AA and Marmet, L and Bernard, JE},
  journal={Applied Physics B},
  volume={67},
  pages={229--234},
  year={1998},
  publisher={Springer},
  doi={https://doi.org/10.1007/s003400050498}
}

@article{olson2014tunable,
  title={Tunable Landau-Zener transitions in a spin-orbit-coupled {Bose}-{Einstein} condensate},
  author={Olson, Abraham J and Wang, Su-Ju and Niffenegger, Robert J and Li, Chuan-Hsun and Greene, Chris H and Chen, Yong P},
  journal={Physical Review A},
  volume={90},
  number={1},
  pages={013616},
  year={2014},
  publisher={APS},
  doi={https://doi.org/10.1103/PhysRevA.90.013616}
}

@article{li2019spin,
  title={Spin current generation and relaxation in a quenched spin-orbit-coupled {Bose}-{Einstein} condensate},
  author={Li, Chuan-Hsun and Qu, Chunlei and Niffenegger, Robert J and Wang, Su-Ju and He, Mingyuan and Blasing, David B and Olson, Abraham J and Greene, Chris H and Lyanda-Geller, Yuli and Zhou, Qi and others},
  journal={Nature Communications},
  volume={10},
  number={1},
  pages={375},
  year={2019},
  publisher={Nature Publishing Group UK London},
  doi={https://doi.org/10.1038/s41467-018-08119-4}
}

@article{ruster2016long,
  title={A long-lived Zeeman trapped-ion qubit},
  author={Ruster, Thomas and Schmiegelow, Christian T and Kaufmann, Henning and Warschburger, Claudia and Schmidt-Kaler, Ferdinand and Poschinger, Ulrich G},
  journal={Applied Physics B},
  volume={122},
  number={10},
  pages={254},
  year={2016},
  publisher={Springer},
  doi={https://doi.org/10.1007/s00340-016-6527-4}
}

@article{ozeri2007errors,
  title={Errors in trapped-ion quantum gates due to spontaneous photon scattering},
  author={Ozeri, Roee and Itano, Wayne M and Blakestad, RB and Britton, Joseph and Chiaverini, J and Jost, John D and Langer, C and Leibfried, Dietrich and Reichle, Rainer and Seidelin, Signe and others},
  journal={Physical Review A—Atomic, Molecular, and Optical Physics},
  volume={75},
  number={4},
  pages={042329},
  year={2007},
  publisher={APS},
  doi={https://doi.org/10.1103/PhysRevA.75.042329}
}

@article{bluvstein2024logical,
  title={Logical quantum processor based on reconfigurable atom arrays},
  author={Bluvstein, Dolev and Evered, Simon J and Geim, Alexandra A and Li, Sophie H and Zhou, Hengyun and Manovitz, Tom and Ebadi, Sepehr and Cain, Madelyn and Kalinowski, Marcin and Hangleiter, Dominik and others},
  journal={Nature},
  volume={626},
  number={7997},
  pages={58--65},
  year={2024},
  publisher={Nature Publishing Group UK London},
  doi={https://doi.org/10.1038/s41586-023-06927-3}
}

@article{chauhan_trapped_2024,
  title = {Chip scale coil stabilized Brillouin laser driving a room temperature trapped ion qubit},
  author = {Chauhan, Nitesh and Caron, Christopher and Wang, Jiawei and Isichenko, Andrei and Helaly, Nishat and Liu, Kaikai and Niffenegger, Robert J. and Blumenthal, Daniel J.},
  journal={Nature Communications},
  volume={17},
  number={3982},
  year={2026},
  publisher={Nature Publishing Group UK London},
  doi={https://doi.org/10.1038/s41467-026-69948-2}
}

\newpage

\onecolumngrid

\newpage

\section{Supplementary Information}

\subsection{Design Conventions}

Following some conventions (i.e. design rules) helps organize layouts, avoiding complex routing that is difficult to verify and debug. 

Conventions we have found useful include:

\begin{enumerate}
    \item \textbf{All beams at a fixed height.}
    \item \textbf{All beams along a fixed grid.}
    This ensures that all baseplates are compatible and easily aligned with each other.
    This also eases the conception of optical layout upon fixed `cardinal directions'.
    \item \textbf{One way branching.}
    All baseplates branch off the same direction from the main beam (all left or all right) so all outputs into optical fibers are the same direction.
\end{enumerate}

These conventions are meant to be broken for specific cases, but setting conventions aids in the modularization of larger systems.

\subsection{Single-pass AOM Baseplate}
Acoustic optical modulators (AOMs) are a core control element within most AMO apparatus, as they enable precise and fast control of laser beam intensity and frequency.
Here we start with a simple design for a single-pass AOM baseplate which branches off of the main beam line from the ECDL. 

Two optional input mirrors (BB05-E02, KM05) allow precise downstream alignment if necessary. Then rotation of a half waveplate (in a RSP05 mount) allows a polarizing beam splitter (PBS101) to branch off a controlled amount of optical power to the AOM. 
Diverted power enters the AOM (Isomet M1250-T200L-0.5) and is diffracted when RF power is pulsed through the AOM. 
Two mirrors allow fine adjustment of the alignment into the fiber port (PAF2-A4A), directing light through the optical fiber to the ion trap.


\subsection{Double-pass AOM Baseplate}

Double-pass AOM configurations are a common method for precise frequency and amplitude control over hundreds of MHz and are common in AMO experiments. 
Similar to the single-pass AOM configuration, two optional input mirrors allow realignment of the input beam and a half waveplate allows precise control of how much power is sent into the baseplate from the PBS and how much transmits to subsequent base plates. Two input mirrors after the PBS enable independent alignment into the AOM (Isomet). 
Unlike the single-pass AOM configuration, a lens is placed a focal length away from the AOM so all diffracted beams are parallel to the input beam. Upon retro-reflection by a mirror, a 2nd `pass' through the AOM then exactly cancels out the diffraction angle of the 1st pass so that the output laser beam counter-propagates with the original input beam, independent of the frequency sent to the AOM. This output beam now has precisely twice the frequency added from each pass through the AOM.
The lens in the retro-reflection path also focuses the laser beamwaists, so the retro-reflection mirror must be placed a focal length away from the lens to ensure the output beam is collimated. 

To prevent the input light from leaking into the output, an iris blocks the zeroth-order beam (and unwanted diffraction orders) just before the retro-reflection mirror, near the beam's focus, but leaving space for a power meter probe.
For our application, the rubidium transition we are locking to is 440 MHz red detuned from the strontium ion transition, so the angle of the AOM and the iris are aligned to transmit the `+1' diffraction order (diffracted \textit{away} from the RF input port). The design can also be reconfigured for the `-1' order by sliding the iris toward the AOM RF input port and optimizing the AOM angle. 
To enable spatial separation of the counter-propagating output beam, a quarter waveplate is inserted into the retro-reflection path to rotate the input vertical polarization to horizontal so that the output beam transmits the initial PBS. Lastly, two mirrors are used to couple light into the optical fiber and to the trapped ion.





\subsection{Extended Cavity Diode Laser}

Homemade extended cavity diode lasers at near-IR wavelengths were an enabling tool for pioneering AMO labs \cite{arnold1998simple} which needed many narrow linewidth lasers for various atomic transitions and operations such as laser cooling and atomic state preparation and/or repumping. Not only were they completely customizable but were often an order of magnitude cheaper than commercial lasers (if commercial options existed at all). Also, the performance of these homemade ECDLs, including specifications such as linewidth and stability were often sufficient for these niche applications, which often do not require broad mode-hop-free scan ranges. However, these designs often require machining multiple custom components \cite{arnold1998simple, hawthorn2001littrow,cook2012high, dutta2012mode, doret2018simple, chang2023low, schkolnik2019extended, daffurn2021simple, kurbis2020extended}, and are static designs for a single wavelength.

Here we focus on designing an extremely simple ECDL with only a single custom part, a dynamic 3D printed mount for the diffraction grating. All other parts are off-the-shelf from commercial vendors with no modification or machining at all. The 3D printed grating mount attaches directly to the ubiquitous KM100PM Thorlabs mount without modification. As a test case we demonstrate laser operation in the violet, specifically 421.6 nm which is useful for strontium trapped ions and also neutral rubidium for Rydberg applications. Until relatively recently, second-harmonic generation was required to achieve these wavelengths with sufficient power for various AMO applications like laser cooling.

\subsubsection{Laser Diode}
The laser diode is made by TopGaN (and Nichia) and lases directly in the violet without second harmonic generation. It is not AR coated and is mounted in 1/2 inch optics, S05LM56 and LMR05. As shown in Figure \ref{fig:ECDL}, the light is collimated with a C610TMD-A lens (focal length f=4mm) mounted within a S05TM09 adapter and a 1/2 inch lens tube (SM05L05) threaded to the front of the LMR05. The LMR05 is mounted to a KM100PM which is then mounted onto a small aluminum block, which does not require any machining and can be cut to an arbitrary size by hand from 1/4 inch thick aluminum stock with a single drilled and threaded hole to mount the KM100PM. This aluminum block is epoxied to a TEC (TECH4), which is then epoxied onto a slightly larger 1/4 inch thick aluminum block (which also does not require machining). The TEC is controlled with a temperature controller such as the Thorlabs TED200C or Koheron CTL200-0. Finally, the bottom plate can be mounted directly to the optical table for thermal sinking and directly match the height of the baseplates. Alternatively, the bottom plate can be placed on a sorbothane sheet for vibration isolation. The full mount is then enclosed in two 3D printed boxes for thermal isolation and stability, as well as some acoustic shielding (with additional layers improving the isolation).

\subsubsection{Dynamic Grating Mount}
The feedback for the extended cavity diode laser is provided by a diffraction grating (GH13-36U) mounted on a parameterized custom 3D printed mount. The Littrow angle of the grating in the mount is set within the code by inputting the wavelength and the diffraction grating pitch, allowing the design to be dynamically reconfigured for arbitrary wavelengths (and AR vs. non-AR diodes etc.). The diffraction grating is paired with a parallel mirror (ME05S-P01) to offset any angles during adjustment and mounted to the front of the KM100PM also holding the laser diode. A PZT element (PA4HKW) is placed behind the grating for precise tuning. While more sophisticated mounts are possible which only tune the angle of the grating through careful placement of the pivot point, they are not readily machined from aluminum. We anticipate the investigation of the stability of sophisticated 3D printed mounts with different materials for these applications to be an interesting direction of future research. The present design is as simple as possible to minimize machining. However, the 3D printed mount shows remarkable stability (see SI) and was used for all experiments.


\subsubsection{Temperature Control}

Laser diodes are never at the exact wavelength required to address a specific atomic transition, and pre-selection of diodes to within a nanometer can easily quadruple their cost. Therefore, it is advantageous to be able to widely tune the laser diode wavelength by more than a nanometer.
Coarse adjustment of the angle of the diffraction grating can broadly tune the laser diode frequency hundreds of GHz. However, tuning the temperature of the laser diode can tune the frequency multiple THz, especially if the diode can be heated. Tuning via cooling is limited due to the risk of condensation, but laser diodes can often be heated more than 30 degrees, to above 50C, at the cost of decreased optical power from the increased current threshold for lasing. 

We tested multiple laser diodes and were able to tune a laser diode which had a free-running wavelength of 418.9 nm up to 421.6 nm (-4.6 THz) by heating the mount up to 50 C. Even at these high temperatures, the thermal stability of the laser diode was within 100 MHz over the course of hours measured by a MOGLabs Fizeau wavemeter and verified with saturated absorption spectroscopy (as detailed below). For these high current cases we used the Thorlabs temperature controller TED200C but for other cases that did not require as much power we used the Koheron CTL200-0 due to the convenient GUI control. 


\subsubsection{Optical Isolator and Cylindrical Telescope}
Next two mirrors are used to align the laser through an optical isolator (IO-3D-405-PBS). 
The efficiency of the transmission through the optical isolator is limited by the ellipticity of the laser beam, which causes clipping.
Therefore, we implement a cylindrical telescope (LJ1821L1-A and LK1085L1-A) to obtain a circular beam shape. Each lens was mounted in a custom 3D printed mount. Alternatively, anamorphic prism pairs could be custom mounted within a 3D printed mount to set the required relative angle. 


\subsection{3D Printing Procedures and Best Practices}

All prints used 50\textmu m layer height with the appropriate exposure time recommended by the resin manufacturer. For printing baseplates, we found it best to hollow the print with 1.5mm wall thickness and 40\% density 3D grid infill pattern. The prints were angled to prevent the high stress of printing large flat faces from pulling the edges of the print off of the build plate. We found an angle of 15 degrees worked best. Because of this, slicer-generated support structures were also used to help support the underside of the baseplates. This tilt should be should be done such that the bottom of the baseplate is angled towards the build plate, ensuring that the supports do not create defects on the top surface. Drainage holes of 3mm diameter were added to the surface of the print to prevent resin from pooling inside. For small adapters and sub-mounts, solid printing was preferable and angling of the prints was not required.

\subsection{Laser Comparison}

We have used a commercial laser and our custom laser for trapped ion cooling and detection interchangeably in our laboratory. The commercial laser supplies 70 mW of power after the optical isolator and our custom PyOpticL ECDL supplies 60 mW of optical power. The commercial laser costs \$35k and has a laser linewidth of 53(1) kHz. The PyOpticL laser has a linewidth of 54(1) kHz and costs just under \$7k, with almost half of the cost being the relatively expensive laser diode.  
We again note that if cost is paramount, the PZT controller can also be made custom for a tenth of the cost of commercial controllers \cite{doret2018simple}. The linewidths are measured with a custom optical frequency discriminator (OFD) comprised of an unbalanced Mach Zehnder interferometer measured using a high bandwidth balanced photodiode and oscilloscope. The frequency noise is then used to calculate the reverse integral linewidth (RIL) of each laser. As with the RIL, we find that the fundamental linewidth of each laser is also similar, both are about 6 kHz.

\begin{figure*}[]
\centering
\includegraphics[width=0.85\textwidth]{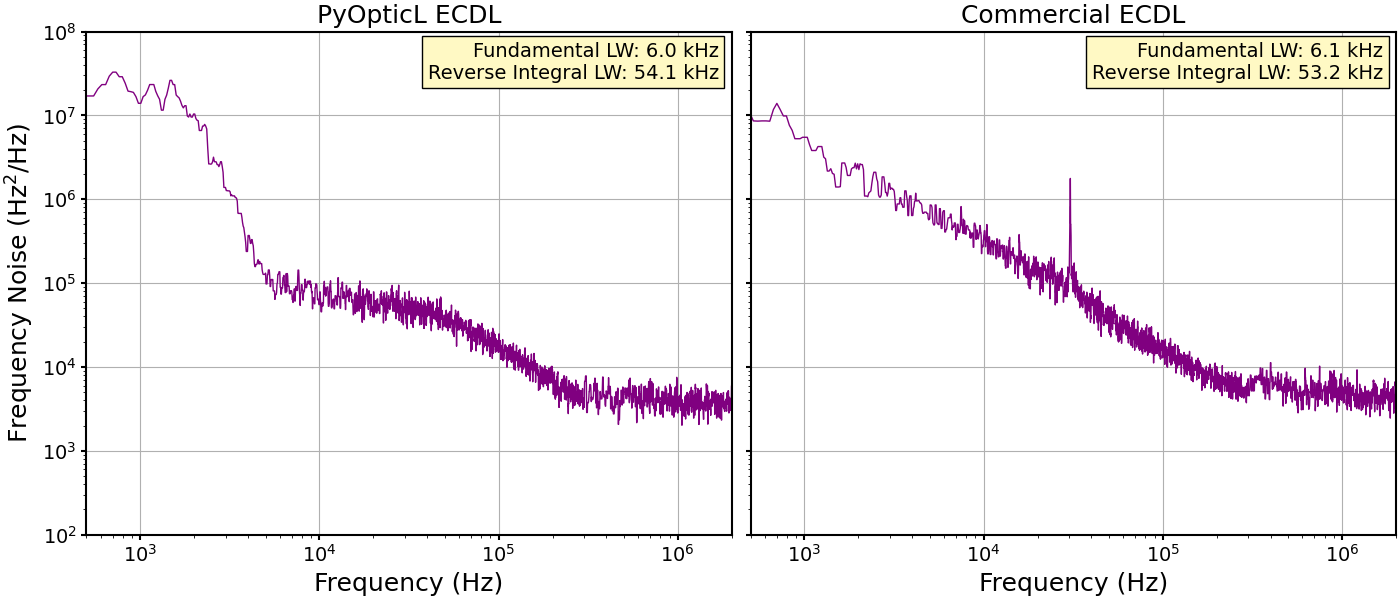}
\caption{Comparison of frequency noise and linewidth of (a) PyOpticL laser and (b) commercial laser. They have very similar performance as measured by a custom optical frequency discriminator (OFD), with both exhibiting about 50 kHz linewidths. }
\label{fig:linewidth}
\end{figure*}

\begin{table}
\centering

\begin{tabular}{| l | l | l |}
\hline
 & \textbf{Commercial ECDL} & \textbf{PyOpticL ECDL} \\
\hline
Power after optical isolator & 70 mW & 60 mW \\
\hline
Linewidth & 53(1) kHz & 54(1) kHz \\
\hline
\textbf{Total cost} & \textbf{\$35,000} & \textbf{\$6,806} \\
\hline

\end{tabular}
\caption{{Comparison of commercial and PyOpticL laser costs}}
\end{table}

\begin{table}
\centering

\begin{tabular}{| l | l |}
\hline
\textbf{Laser Component} & \textbf{Cost} \\
\hline
Diode  & \$3000 \\
\hline
Temperature/Current controller & \$1140 \\
\hline
PZT controller & \$900 \\
\hline
Optical Isolator & \$1170 \\
\hline
Thorlabs parts & \$596 \\
\hline
\textbf{Total} & \textbf{\$6,806} \\
\hline

\end{tabular}
\caption{Summary of major costs for PyOpticL laser}
\end{table}

\begin{table}[h]
\centering

\begin{tabular}{| l | l | l | l |}
\hline
\textbf{Part Name} & \textbf{Part ID } & \textbf{Cost} & \textbf{Link} \\
\hline
Lens & C610TMD-A & \$155 & \href{https://www.thorlabs.com/item/C610TMD-A}{https://www.thorlabs.com/item/C610TMD-A} \\
\hline
Lens Adapter & S05TM09 & \$23 & \href{https://www.thorlabs.com/item/S05TM09}{https://www.thorlabs.com/item/S05TM09} \\
\hline
Lens Tube  & SM05L05 & \$16 & \href{https://www.thorlabs.com/item/SM05L05}{https://www.thorlabs.com/item/SM05L05} \\
\hline
Laser diode adapter & S05LM56 & \$32 & \href{https://www.thorlabs.com/item/S05LM56}{https://www.thorlabs.com/item/S05LM56} \\
\hline
Laser diode mount & LMR05 & \$18 & \href{https://www.thorlabs.com/item/LMR05}{https://www.thorlabs.com/item/LMR05} \\
\hline
Mount & KM100PM & \$94 & \href{https://www.thorlabs.com/item/KM100PM}{https://www.thorlabs.com/item/KM100PM} \\
\hline
Grating & GH13-36U & \$106 & \href{https://www.thorlabs.com/item/GH13-36U}{https://www.thorlabs.com/item/GH13-36U} \\
\hline
Parallel Mirror & ME05S-P01 & \$26 & \href{https://www.thorlabs.com/item/ME05S-P01}{https://www.thorlabs.com/item/ME05S-P01} \\
\hline
PZT & PA4HKW & \$101 & \href{https://www.thorlabs.com/item/PA4HKW}{https://www.thorlabs.com/item/PA4HKW} \\
\hline
TEC & TECH4 & \$25 & \href{https://www.thorlabs.com/item/TECH4}{https://www.thorlabs.com/item/TECH4} \\
\hline
\textbf{TOTAL COST} &  & \textbf{\$596} &  \\
\hline

\end{tabular}
\caption{List of all minor components in the PyOpticL laser.}
\end{table}

\begin{figure*}[]
\centering
\includegraphics[width=0.9\textwidth]{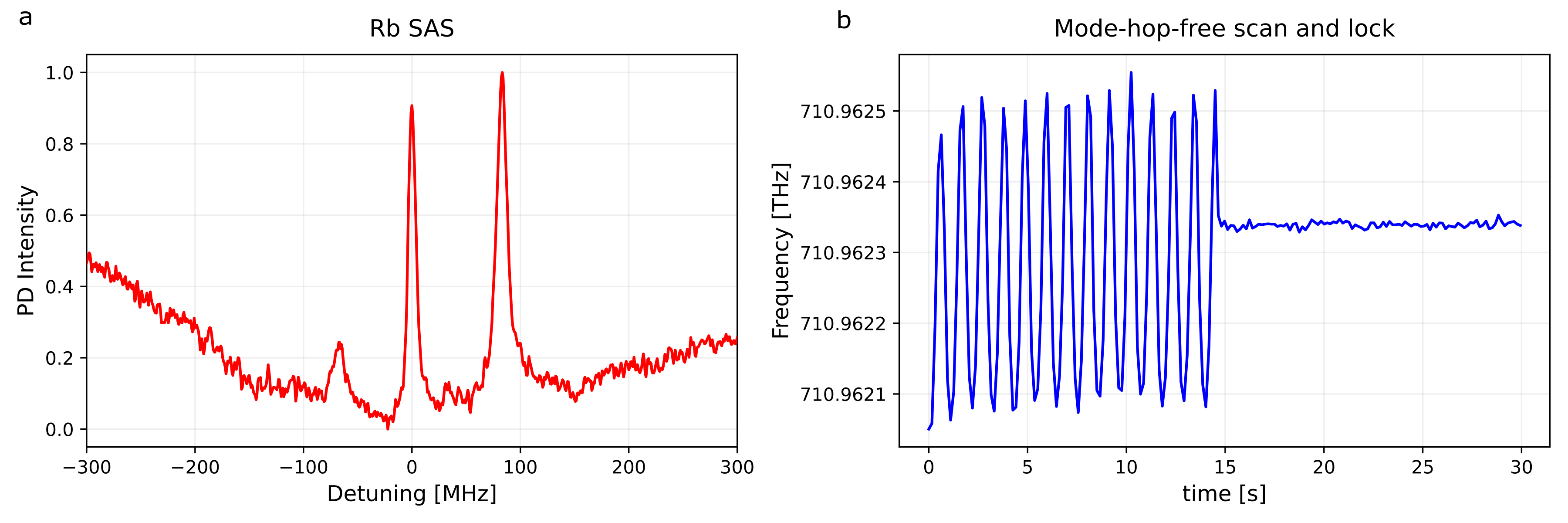}
\caption{ 
\textbf{Saturated Absorption Spectroscopy:}
a) Doppler free spectroscopy of the $5\text{s}^2\text{S}_{1/2} \rightarrow 6\text{p}^2\text{P}_{1/2}$ transitions of $^{85}\text{Rb}$  (F = 2 $\rightarrow$ 2,3) near 421.6 nm. b) Slowly scanning the PZT voltage shows mode-hop-free tuning over 400MHz. The laser is then locked to the center peak for absolute stabilization. 
}
\label{fig:SAS}
\end{figure*}

\subsection{Saturated Absorption Spectroscopy}

To demonstrate stable single-mode operation of the ECDL we use it to observe saturated absorption spectroscopy (SAS). SAS is commonly used to lock and stabilize lasers using an atomic transition. We note that \textit{many} other examples of small SAS setups exist \cite{knappe2007microfabricated, kurbis2020extended, madkhaly2021performance, christ2024additive} and are commercially available \cite{sas_thorlabs,sas_vescent}.


Here we create the entire saturated absorption spectroscopy (SAS) layout using PyOpticL (Fig. \ref{fig:dynamic}), and build the 1/2 inch mounted optics plate (Fig. \ref{fig:SAS}b). 

Similar to other baseplates, two optional input mirrors allow downstream alignment and a half waveplate allows power to be branched off with a polarizing beam splitter. A beam sampler then reflects a small amount of light to traverse the Rb cell as a `probe' beam. A half-waveplate rotates the vertical polarization to horizontal so it can transmit through the final PBS and be detected by the photodiode. The rest of the optical power transmits the beam sampler as the `pump' beam, with two mirrors for alignment and a PBS to allow counter propagation. The pump beam saturates the excitation of the vapor atoms at zero velocity relative to the pump for Doppler free spectroscopy. The pump power can optionally be reused by fiber coupling to a wavemeter.

The frequency was tuned by applying a high voltage (up to 150V) to the PZT and enabled mode hop free tuning over 1 GHz. Synchronous ramping with the current controller could allow broader mode hop free tuning \cite{dutta2012mode}. This allowed the laser to be tuned to the resonance of rubidium and enabled capture of Doppler free peaks as shown in Figure \ref{fig:SAS}a. 


\subsection{Alignment and Power Stability}
\begin{figure*}[]
\centering
\includegraphics[width=0.9\textwidth]{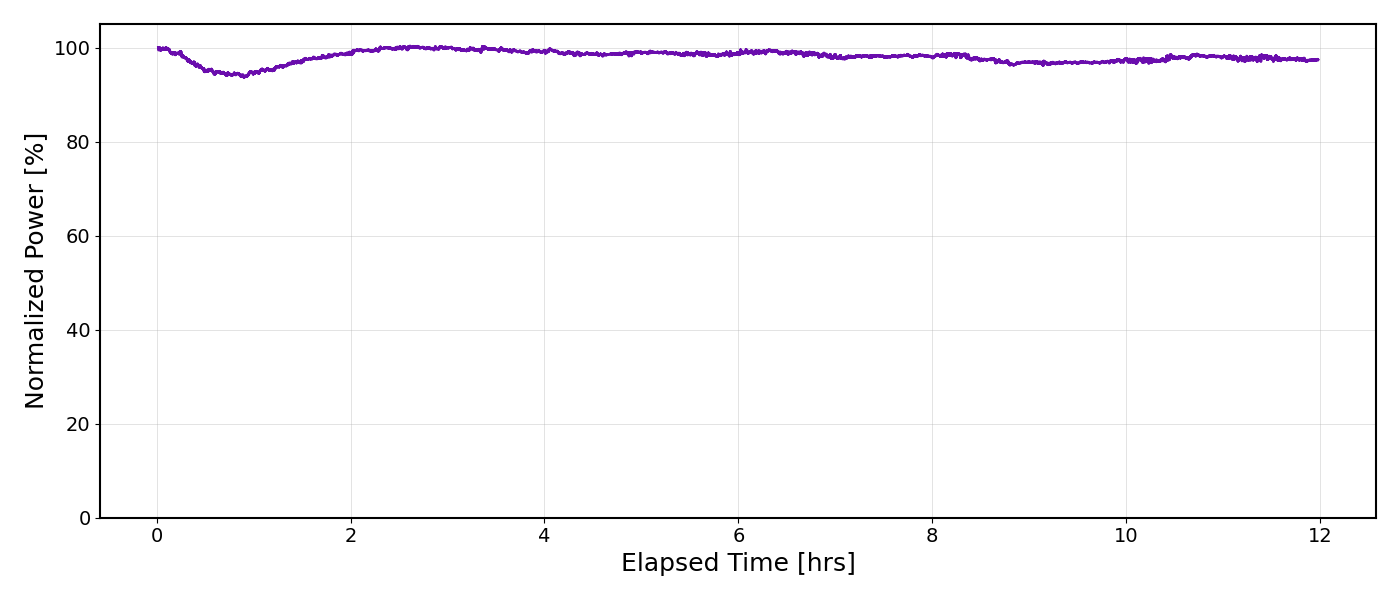}
\caption{ 
\textbf{Power Stability Measurement:}
We benchmark the alignment stability of the Doppler Subsystem by observing the power coupled to a fiber over a 12 hour period. After an initial period post-alignment, the power settles to 98\% of its initial value.}
\label{fig:PowerStability}
\end{figure*}

We measure the alignment stability of a PyOpticL subsystem by observing the coupled power through a fiberport and plotting in Fig. \ref{fig:PowerStability}. Here, we are running a commercial ECDL laser at 422 nm through our `Doppler Cooling' subsystem for strontium ions. The subsystem is comprised of a mix of 3D printed and CNC machined plates and includes a periscope, telescope, modulation transfer spectroscopy (MTS) baseplate, and 2 doublepass baseplates (in our setup, the MTS baseplate and first doublepass baseplate are 3D printed). We allow 0th order light from the second doublepass to be fiber coupled into a ThorLabs PAF2-5A fiberport and observe the coupled power over a 12 hour period. After adjusting the alignment, we observe a settling period in which the coupled power fluctuates before finally converging to 98\% of the original coupled power.

\subsection{Dynamic Saturated Absorption Spectroscopy Baseplate}



\begin{lstlisting}[language=Python, style=myPythonStyle, basicstyle=\footnotesize, frame=single]
from PyOpticL.beam_path import BeamPath
from PyOpticL.layout import Component
from PyOpticL.library import baseplate
from PyOpticL.utils import Dimension as dim
from PyOpticL.utils import cardinal_angle, turn_angle, fix_relative_imports

fix_relative_imports()

from parameters import get_scale_parameters

scale_params = get_scale_parameters("half_inch_mounted")

rb_sas_baseplate = Component(
    label="Rb SAS",
    definition=baseplate(
        dimensions=(
            dim(18, "in") * scale_params["overall_scale"],
            dim(6, "in") * scale_params["overall_scale"],
            scale_params["baseplate_height"],
        ),
        optical_height=scale_params["optical_height"],
    ),
)

beam = rb_sas_baseplate.add(BeamPath(label="Beam", wavelength=780, waist=scale_params["beam_waist"]),  position=(dim(15.5, "grid") * scale_params["overall_scale"], 0, 0), rotation=(0, 0, cardinal_angle["up"]))

beam.add(Component(label="Input Mirror 1", definition=scale_params["mirror"]), beam_index=0b1, distance=dim(1.5, "in") * scale_params["overall_scale"], rotation=(0, 0, turn_angle["up-right"]))

beam.add(Component(label="Input Mirror 2", definition=scale_params["mirror"]), beam_index=0b1, distance=dim(1, "grid") * scale_params["overall_scale"], rotation=(0, 0, turn_angle["right-up"]))

beam.add(Component(label="Input Half Waveplate", definition=scale_params["waveplate"]), beam_index=0b1, distance=dim(1.5, "in") * scale_params["overall_scale"], rotation=(0, 0, cardinal_angle["up"]))

beam.add(Component(label="Beam Splitter 1", definition=scale_params["beamsplitter"]), beam_index=0b1, distance=dim(2, "in") * scale_params["overall_scale"], rotation=(0, 0, cardinal_angle["up"]))


beam.add(Component(label="Input Mirror 3", definition=scale_params["mirror"]), beam_index=0b11, distance=dim(3.5, "in") * scale_params["overall_scale"], rotation=(0, 0, turn_angle["left-down"]))

beam.add(Component(label="Pump/Probe Splitter", definition=scale_params["circular_sampler"]), beam_index=0b11, distance=dim(0.75, "in") * scale_params["overall_scale"], rotation=(0, 0, turn_angle["down-left"]))

beam.add(Component(label="Probe Half Waveplate", definition=scale_params["waveplate"]), beam_index=0b111, distance=dim(1.75, "in") * scale_params["overall_scale"], rotation=(0, 0, cardinal_angle["left"]))

beam.add(Component(label="Probe Mirror 1", definition=scale_params["mirror"]), beam_index=0b111, distance=dim(1.25, "in") * scale_params["overall_scale"], rotation=(0, 0, turn_angle["left-down"]))

beam.add(Component(label="Probe Mirror 2", definition=scale_params["mirror"]), beam_index=0b111, y_position=dim(3, "in") * scale_params["overall_scale"], rotation=(0, 0, turn_angle["down-left"]))

beam.add(Component(label="Rb Cell", definition=scale_params["rb_cell_definition"]), beam_index=0b111, distance=dim(3.5, "in") * scale_params["overall_scale"], rotation=(0, 0, cardinal_angle["right"]))

beam.add(Component(label="Pump Mirror 1", definition=scale_params["mirror"]), beam_index=0b110, distance=dim(3, "in") * scale_params["overall_scale"], rotation=(0, 0, turn_angle["down-left"]))

beam.add(Component(label="Pump Half Waveplate", definition=scale_params["waveplate"]), beam_index=0b110, distance=dim(4, "in") * scale_params["overall_scale"], rotation=(0, 0, cardinal_angle["left"]))

beam.add(Component(label="Pump Mirror 2", definition=scale_params["mirror"]), beam_index=0b110, distance=dim(5.5, "in") * scale_params["overall_scale"], rotation=(0, 0, turn_angle["left-up"]))

beam.add(Component(label="Beam Splitter 2", definition=scale_params["beamsplitter"]), beam_index=0b110, y_position=dim(3, "in") * scale_params["overall_scale"], rotation=(0, 0, cardinal_angle["left"]))

beam.add(Component(label="Photodetector", definition=scale_params["photodetector"]), beam_index=0b1110, **scale_params["photodetector_constraint"], rotation=(0, 0, cardinal_angle["right"]))

if __name__ == "__main__":
    rb_sas_baseplate.recompute()


\end{lstlisting}


\newpage

\twocolumngrid

\end{document}